\documentclass[aps,preprinti, showpacs]{revtex4}%
\usepackage{epsfig}
\usepackage{bm}

\usepackage{amsfonts}
\usepackage{amsmath}
\usepackage{amssymb}
\usepackage{graphicx}%
\begin{document}
\title[Short title for running header]{Three-atom scattering via the Faddeev scheme in configuration space.}

\author{Walter Gl\"ockle}
\affiliation{ Institut f\"ur theoretische Physik II, Ruhr Universit\"at Bochum,
D-44780 Bochum, Germany}
\author{George Rawitscher}
\affiliation{Department of Physics, University of Connecticut, Storrs, CT 06268}
\keywords{one two three}

\date{\today}

\begin{abstract}
 Faddeev equations in configuration space and integral form for three-atom scattering processes are formulated allowing for additive and nonadditive forces. The explicit partial wave
 decomposition is displayed. This formulation appears to be a valuable alternative
 to current approaches based on hyperspherical harmonic expansion methods of the
 Schr\"odinger equation.
\end{abstract}





\pacs{PACS number}

\maketitle

\section{Introduction}

The Faddeev scheme for three bodies is a very successful framework in nuclear
physics and specifically in the 3-nucleon system \cite{gloeckle96},\cite{golak05}.
 Most of
the calculations are performed in momentum space, which is naturally adapted
to nuclear forces derived from meson exchange diagrams. Also the
coordinate space version of the Faddev equations in differential form has been
successfully used, especially for the three-nucleon bound states \cite{chen86}, but an
integral equation  form of the Faddeev equation in configuration space has not yet
 been developed in detail. To the best
of our knowledge this has not been done before, neither in the fields of  nuclear nor
atomic physics. 

While in nuclear physics there is only one 2-body bound state, the deuteron,
diatomic  molecules have numerous vibrational and rotational levels. This poses
quite a numerical challenge, especially in 3-atom scattering processes. A
great deal of work has been done in this field using hyperspherical harmonic
expansion methods,  \cite{colav},\cite{desh}. Here and below we quote only a few recent papers.
 It is the purpose of this paper  to describe
 an alternative approach
using the configuration space Faddeev equations in integral form. 
 We have chosen that scheme since over the years a  highly efficient
numerical technique in coordinate space has been developed by one of us \cite{Cise}
for solving 2-body scattering and bound state problems. This method is  based on
spectral expansion techniques in terms of Chebyshev polynomialsi denoted as S-IEM in \cite{Cise}.
 As we shall indicate these methods are
ideally suited to  the integral  Faddeev scheme for 3 atoms.
It is hoped that this technique will result in an accuracy for the 
final 3-body results that is higher than what is currently achieved
 in nuclear physics calculations in momentum space ( about
 3 significant figures). Needless to
say there exist a great demand for theoretical support for various kinds of
3-atom processes: 3-atom recombination processes \cite{suno},
collisional cooling of co-trapped atomic and molecular ions by
 ultracold atoms\cite{bodo},\cite{maka},\cite{stw},\cite{bah}, quantum dynamics of ultra cold 
atom-diatom collisions \cite{sold}. Three-atom potential surfaces
 have been developed, for instance for $H3$\cite{ booth} or for Li3 \cite{cvit}.

In section II we provide a brief  derivation of the form of
Faddeev equations we want to use. The coordinate space representation
expressed in terms of vectors is subsequently set up in section III, while its partial wave
 representation is worked out in section IV. Because of the complexity of the
 resulting expressions for general orbital angular momenta a separate section V
 is devoted to an overall s-wave reduction, since this case provides a more transparent
 insight into the structure of the equations.  Technical
details for the partial wave decomposition are deferred to the Appendices.
We delegate  a numerical  feasibility study to a forthcoming paper,
 but we  nevertheless   add some remarks on that issue in section VI.
 Finally  in section VII  we provide a summary and conclusions.
 \medskip

\section{The theoretical framework}

Let us regard a system of three distinguishable atoms, which interact by 2-and
3- atom forces. Though apparently most often  the full potential
surface for 3 atoms is used ab initio, we prefer to separate out
  the 2-atom forces from the
full \ potential surface and treat the remainder as a genuine 3-atom force. Thus we
denote the total potential energy between 3 atoms by%
\begin{eqnarray}
V=V_{1}+V_{2}+V_{3}+V_{4}\label{1}
\end{eqnarray}
where the 2- atom forces are conveniently denoted by $V_{i}\equiv V_{jk}$
$($with $j,k\neq i)$ and $V_{4}$ is the genuine 3- atom force. Instead of
introducing four Faddeev components corresponding to the four contributions to
the potential we split the 3- atom force into 3 parts, which are then combined with
the three 2-atom forces, as follows. 
We introduce the notation
\begin{eqnarray}
V_{4}=V_{4}^{(1)}+V_{4}^{(2)}+V_{4}^{(3)}\label{2}
\end{eqnarray}
where $V_{4}^{(i)},$ $i=1,2,3$\ has the property of beeing composed of two
pieces, one symmetrical and the other antisymmetrical in the interchange of 
 atoms $j,k\neq
i$. In detail

\begin{eqnarray}
V_{4}  &  \equiv &  V(1,2,3)=\frac{1}{6}[V(123)+V(132)]+\frac{1}{6}[V(123)-V(132)]\label{3}\\
&  + &  \frac{1}{6}[V(123)+V(321)]+\frac{1}{6}[V(123)-V(321)]\nonumber\\
&  + & \frac{1}{6}[V(123)+V(213)]+\frac{1}{6}[V(123)-V(213)]\nonumber
\end{eqnarray}
where the  argument 123 stands for the space coordinate vectors of the 3 atoms. Thus
the two terms in the first row are symmetrical and antisymmetrical, respectively, under
exchange of atoms 2 and 3 and similarly for the two terms in next two rows.
We group the potential into three terms $V_{i}$ and $V_{4}^{(i)}$ in
 the 3-atom Schr\"odinger
equation (SE), which then can be written as
\begin{equation}
(H_{0}-E)\Psi=-\sum_{i=1}^{3}(V_{i}+V_{4}^{(i)})\Psi. \label{4}
\end{equation}

In the present paper we describe a scattering process where one atom is incident on a molecular bound
state of the other two atoms. We can number the three atoms such that the
target molecule is formed out of atoms 2 and 3. Thus the initial channel state
is
\begin{eqnarray}
\vert \phi_{1}> =  |u>_{23}\vert \vec{q}_{0}>_{1} \label{5}
\end{eqnarray}
composed of the 2-atom bound state $\ |u>$ with quantum numbers to be
specified later and a momentum eigenstate $|\vec{q}_{0}>_{1}$ of relative
motion of atom 1 and the molecule (2,3). That channel state obeys 
\begin{eqnarray}
(H_{0}+V_{1})|\phi_{1}>=E\ |\phi_{1}>\label{6}
\end{eqnarray}
\ where $E$ is the total center of mass energy, and $H_{0}$ is the kinetic
energy operator for the three atoms.

The integral form of the SE is
\begin{eqnarray}
\Psi=G_{0}\left(  \sum_{i=1}^{3}(V_{i}+V_{4}^{(i)})\Psi\right)\label{7}
\end{eqnarray}
where  where $G_{0}$ is the free 3-atom propagator, $G_{0}=(E+i\varepsilon
-H_{0})^{-1}$. There is no extra driving term, since there is no solution
  to the left hand side of Eq.(\ref{4})
alone which is regular at the origin and purely outgoing. 
The above equation  suggests the decomposition
\begin{equation}
\Psi\equiv\sum_{i=1}^{3}\psi_{i}\label{8}
\end{equation}
with
\begin{equation}
\psi_{i}\equiv G_{0}(V_{i}+V_{4}^{(i)})\Psi\label{9}
\end{equation}
Inserting the decomposition (\ref{8}) into the right hand side of
Eq.(\ref{9}) and using the well known  identities 
\begin{eqnarray}
( 1 - G_0 V_1 )^{(-1)} G_0 V_1 = G_0 t_1 \label{10}\\
( 1-  G_0 V_1 )^{(-1)} = 1 + G_0 t_1 \label{11}
\end{eqnarray} 
following standard steps \cite{gloeckle83} \cite{hueber97}  one ends up with the
 3 coupled
Faddeev equations(FE)
\begin{eqnarray}
\psi_{1}  &  =\phi_{1}+G_{0}\ t_{1}(\psi_{2}+\psi_{3})+(1+G_{0}t_{1}
)G_{0}\ V_{4}^{(1)}\Psi\nonumber\\
\psi_{2}  &  =G_{0}\ t_{2}(\psi_{3}+\psi_{1})+(1+G_{0}t_{2})G_{0}\ V_{4}
^{(2)}\Psi\nonumber\\
\psi_{3}  &  =G_{0}\ t_{3}(\psi_{1}+\psi_{2})+(1+G_{0}t_{3})G_{0}\ V_{4}
^{(3)}\Psi \label{12}
\end{eqnarray}
The new ingredient, the 2- atom t-operators $t_{i}$ embedded into the 3-body space,
 with $i=1,2,3,$ obey the
Lippmann Schwinger equation(LSE)
\begin{eqnarray}
t_{i}=V_{i}+V_{i}G_{0}t_{i} \label{13}
\end{eqnarray}

 This generalization  is natural, since the Faddeev equations
contain a two-atom potential $ V_i$  separately in each arrangement $i$. A short
description of the related  two-body operator $ \tau(r,r^{\prime})$ and the amplitude $%
T\left( r\right) $ is as follows:

The Lippmann Schwinger equation for the solution of the two-body  SE
is $\psi=\phi+g_{0}V\psi$, where $g_{0}$ is the Green's function $
(E_{2}+i\varepsilon-H_{02})^{-1}$, $H_{02}$ is the two-body kinetic energy
operator, $V$ the potential, and $\phi$ is the incident plane wave. The
product $V(r)\ \psi(r)$ is denoted by $T(r)$, and is expressed in terms
of an integral involving the $\tau$ operator as
\begin{eqnarray}
V(r)\ \psi(r)=\int_{0}^{\infty}\tau(r,r^{\prime})\ \phi(r^{\prime})\
dr' r'^2 \label{14}. 
\end{eqnarray}
By manipulating operator identities ( \ref{10}) and (\ref{11}) 
one can show that $(1-Vg_{0})^{-1}V=\tau$,
which is equivalent to Eq. (\ref{13}), with the difference that $g_{0}$
contains only the kinetic energy for the relative motion of two particles,
while $G_{0}$ also includes the kinetic energy for the  other particle,
and further, $E_{2}$ is the two-body energy, while $E$ is the total
three-body energy. The connection between $t_{i}$ and the two-body $t-$%
matrix $\tau$ is established in section III, see Eqs (\ref{59}) and (\ref{60}).

The kernel pieces in Eqs.(\ref{12}) determine the asymptotic behavior of the wave function components $\psi_i$ in
the outgoing channels. For the 3- atom break- up they can be read off from the
expressions standing  to the right of the free propagator $G_{0}$ \cite{gloeckle96} 
\cite{gloeckle83}.

Thus if we define the three T-amplitudes as 
\begin{eqnarray}
\psi_{1}  &  =\phi_{1}+G_{0}\ T_{1}\nonumber\\
\psi_{2}  &  =G_{0}\ T_{2}\nonumber\\
\psi_{3}  &  =G_{0}\ T_{3} \label{15}
\end{eqnarray}
then we obtain for the $ T_i$  's the expressions
\begin{eqnarray}
T_{1}=t_{1}G_{0}(T_{2}+T_{3})+(1+t_{1}G_{0})G_{0}\ V_{4}^{(1)}\Psi \label{16}
\end{eqnarray}
and correspondingly for $T_{2}$ and $T_{3}$.  The $T$- amplitudes are more useful
 than the wave function amplitudes because from the $ T $- amplitudes one can derive
 the asymptotic normalization of the wave functions and furthermore , most importantly,
 the $ T$'s decrease to zero at large distances. Inserting Eqs.(\ref{15}) into
the right hand sides of Eq.(\ref{16}) and the ones for $ T_2$ and $ T_3$ 
one finds a set of 3 coupled equations
\begin{eqnarray}
T_{1}=t_{1} G_0 ( T_2 +T_3 )+(1+t_{1}G_{0})\ V_{4}^{(1)}[\phi_{1}
+G_{0}\ (T_{1}+T_{2}+T_{3})] \label{17}
\end{eqnarray}
\begin{eqnarray}
T_{2}  &  = & t_{2}\phi_{1}+(1+t_{2}G_{0})\ V_{4}^{(2)}[\phi_{1}+t_{2}G_{0}
(T_{3}+T_{1})]\nonumber\\
&  + & (1+t_{2}G_{0})V_{4}^{(2)}G_{0}(T_{1}+T_{2}+T_{3}) \label{18}
\end{eqnarray}
\begin{eqnarray}
T_3  &  = & t_3 \phi_1 + (1+t_3 G_0)\ V_{4}^{(3)}[ \phi_1 + t_3 G_0
(T_1+ T_2)]\nonumber\\
&  + & (1+ t_3 G_0 ) V_4^{(3)} G_0 (T_1+ T_2 + T_3 ) \label{19}
\end{eqnarray}
This is the set of equations which are  to be solved. A partial wave
 decomposition will be given in section IV.

Because of (\ref{8}) the total break- up amplitude will be
\begin{eqnarray}
U_0 \equiv T_1+ T_2+ T_3 \label{20}
\end{eqnarray}
from which the physical break- up amplitude can be obtained by calculating the
matrix element
\begin{eqnarray}
<\phi_{0}|U_{0}> \label{21}
\end{eqnarray}
where $|\phi_{0}>$ is a product of momentum eigenstates for the free motion of
the three atoms in the final state.

One is also interested in the transition amplitudes for elastic and
rearrangement scattering. They can  be read off from Eqs.(\ref{12}) by
rewriting the kernel parts such that $ V_i$ distorts  the Green operators
\begin{eqnarray}
G_i \equiv \frac{1}{E+i \varepsilon-H_0-V_i}  \label{22}
\end{eqnarray}
 Since $G_{0}t_{i}=G_{i}V_{i}$ , with $(1+G_{0}t_{i})G_{0}=G_{i},$ one
finds for the elastic transition amplitude
\begin{eqnarray}
U_{11}=V_{1}(\psi_{2}+\psi_{3})+V_{4}^{(1)}\Psi \label{23}
\end{eqnarray}
and for the two rearrangement amplitudes
\begin{eqnarray}
U_{21}    = V_{2}(\psi_{3}+\psi_{1})+V_{4}^{(2)}\Psi \label{24}\\
U_{31}    =V_{3}(\psi_{1}+\psi_{2})+V_{4}^{(3)}\Psi \label{25}
\end{eqnarray}
Upon inserting (\ref{15}) into the above expressions for the $ U$'s, we  obtain
\begin{eqnarray}
U_{11}=V_1 G_0 (T_2 + T_ 3 )+ V_ 4^{(1)}[ \phi_ 1 + G_ 0  (T_ 1 + T_ 2 + T_ 3 )] \label{26}
\end{eqnarray}
\begin{eqnarray}
U_{21}=V_ 2 [ \phi_1 + G_ 0 (T_ 3 + T_ 1 )] + V_ 4^{(2)}[ \phi_ 1 + G_ 0 
 (T_ 1 + T_ 2 + T_ 3 )] \label{27}
\end{eqnarray}
\begin{eqnarray}
U_{31}=V_ 3 [ \phi_ 1 + G_ 0  (T_ 1 + T_ 2 )]+ V_ 4^{(3)}[ \phi_ 1 + G_ 0  (T_ 1 + T_ 2 +
 T_ 3 )] \label{28}
\end{eqnarray}
The physical amplitudes are
\begin{eqnarray}
< \phi_i \vert U_{i1} >  \label{29}
\end{eqnarray}
where the $\phi_{i}$'s obey the SE
\begin{eqnarray}
V_{i}|\phi_{i}>\ =(E-H_{0})|\phi_{i}> \label{30}
\end{eqnarray}
Like in Eq. (\ref{5}) the channel states $\phi_{2}$ and $\phi_{3}$ contain
bound states. Thus when calculating the matrix elements (\ref{29}) one can
replace $V_{i}\phi_{i}$ by $G_{0}^{-1}\phi_{i},$ and hence replace $V_{i}
G_{0}$ in Eqs.(\ref{26}),(\ref{27}),(\ref{28}) \ by $1$. These substitutions then lead to the final
expressions
\begin{eqnarray}
U_{11}=(T_{2}+T_{3})+V_{4}^{(1)}[\phi_{1}+G_{0}\ (T_{1}+T_{2}+T_{3})] \label{31}
\end{eqnarray}
\begin{eqnarray}
U_{21}=G_{0}^{-1}\phi_{1}+(T_{3}+T_{1})+V_{4}^{(2)}[\phi_{1}+G_{0}
\ (T_{1}+T_{2}+T_{3})] \label{32}
\end{eqnarray}
\begin{eqnarray}
U_{31}=G_{0}^{-1}\phi_{1}+(T_{1}+T_{2})]+V_{4}^{(3)}[\phi_{1}+G_{0}
\ (T_{1}+T_{2}+T_{3})] \label{33}
\end{eqnarray}
It can easily be verified that the forms in the equations above for the
$U_{i1}$ are identical to the more standard expressions, which are%
\begin{eqnarray}
U_{i1}=(V_{j}+V_{k}+V_{4})\Psi,~~~j,k\neq i \label{34}
\end{eqnarray}
For instance one has
\begin{eqnarray}
  <\phi_{1}|U_{11}> &  =&  <\phi_{1}|V_{1}(\psi_{2}+\psi_{3})>+<\phi_{1}
|V_4^{(1)}|\Psi> \label{35}\\
&  = &  <\phi_{1}|G_{0}^{-1}(\psi_{2}+\psi_{3})>+<\phi_{1}|V_{4}^{(1)}|\Psi>\nonumber\\
&  =&  <\phi_{1}|G_{0}^{-1}|\Psi>-<\phi_{1}|G_{0}^{-1}|\psi_{1}>+<\phi
_{1}|V_{4}^{(1)}|\Psi>\nonumber\\
&  =& <\phi_{1}|V_{1}+V_{2}+V_{3}+V_{4}|\Psi>-<\phi_{1}|G_{0}^{-1}G_{0}
(V_{1}+V_{4}^{(1)})|\Psi>\nonumber\\
& +  &  <\phi_{1}|V_{4}^{(1)}|\Psi>\nonumber\\
&  = &  <\phi_{1}|V_{2}+V_{3}+V_{4}^{(1)}|\Psi>\nonumber
\end{eqnarray}
In the second to last step we used the definition of the Faddeev amplitudes,
Eq. (\ref{9}).

In the case of identical atoms the 3 Faddeev amplitudes (\ref{9}) are
identical in form, only the particles are permuted \cite{gloeckle83}. One easily finds
\begin{eqnarray}
\psi_{2}+\psi_{3}=\mathcal{P}\psi_{1} \label{36}
\end{eqnarray}
where
\begin{eqnarray}
\mathcal{P\equiv}P_{12}P_{23}+P_{13}P_{21} \label{37}
\end{eqnarray}
\textbf{ }is a sum of a cyclical and anticyclical permutations of 3
objects. Thus the total state can be written as
\begin{eqnarray}
\Psi=(1+\mathcal{P})\psi_{1} \label{38}
\end{eqnarray}
and hence only one FE is needed that  reads
\begin{eqnarray}
\psi_ 1 = \phi_ 1 + G_ 0  t_ 1 \mathcal{P} \psi_ 1 + (1+G_ 0 t_ 1 )G_ 0
 V_ 4^{(1)} (1+\mathcal{P}) \psi_ 1  \label{39}
\end{eqnarray}

?
If one defines again
\begin{eqnarray}
\psi_1= \phi_1 + G_0  T  \label{40}
\end{eqnarray}
then the amplitude $T$ obeys \cite{hueber97}
\begin{eqnarray}
T  &  = & t_ 1 \mathcal{P} \phi_ 1 + (1+t_ 1 G_ 0 )  V_ 4^{(1)} (1+\mathcal{P}) \phi_1 \nonumber\\
& + & t_ 1 \mathcal{P} G_ 0 T   +  (1+ t_1 G_ 0 ) V_ 4^{(1)} (1+\mathcal{P})G_ 0 T \label{41}
\end{eqnarray}
The  driving term in (\ref{41}) is contained in the first line and the integral operator
 is contained in the second line.
The complete break up amplitude is
\begin{eqnarray}
U_ 0 = (1+\mathcal{P})T \label{42}
\end{eqnarray}
Because of the identity of the atoms there is only one amplitude for the
transition into two-body fragmentation, which, according to Eq.(\ref{39})
is
\begin{eqnarray}
U  &  = & V_ 1 \mathcal{P} \psi_ 1 + V_ 4^{(1)} (1+\mathcal{P}) \psi_ 1 \nonumber\\
&  = & \mathcal{P}G_0^{-1} \phi_ 1 + \mathcal{P} T + V_ 4^{(1)} (1+\mathcal{P})
\phi_{1}+ V_ 4^{(1)} (1+\mathcal{P}) G_ 0 T \label{43}
\end{eqnarray}
The last expression is valid for calculating the physical matrix element
$<\phi_{1}|U>.$

This concludes the derivation of the formal framework. The case where only two
atoms are identical is similar to the above and is left to the reader.

\section{Coordinate space representation}

In a 3-atom system there are three 2-body fragmentation ( or arrangement )  channels going with
three types of Jacobi vectors ( ijk = 123 and cyclical permutations)
\begin{align}
\vec{x}^{(i)}    = \vec{x}_{j}- \vec{x}_{k};~~~~\vec{y}^{(i)}=
 \vec{x}_{i}- \frac{m_j}{m_j + m_k} \vec{x}_{j}- \frac{m_k}{ m_j + m_k} 
\vec{x}_{k}
\end{align}
 We introduce
coordinate space states%
\begin{eqnarray}
\vert \vec{x} \ \vec{y}   >_{1} \equiv \vert \vec x^{(1)}, \vec y^{(1)} >\label{45}\\
\vert \vec{x} \ \vec{y}   >_{2} \equiv \vert \vec x^{(2)}, \vec y^{(2)} >\label{46}\\
\vert \vec{x} \ \vec{y}   >_{3} \equiv \vert \vec x^{(3)}, \vec y^{(3)} >\label{47}
\end{eqnarray}
Each set of states is complete:
\begin{eqnarray}
\int d\vec{x}d\vec{y}\ |\vec{x}\vec{y}><\vec{x}\vec{y}|=1.
\end{eqnarray}

The various terms for the $T$-amplitudes in Eqs. (\ref{17}) - ( \ref{19})
 will now be written in the coordinate space representation, using the definitions
 in Eqs. (\ref{45}) - ( \ref{47}).
It is natural to represent  the amplitude $T_{1}$ as $_{1}<\vec{x}\vec{y}|T_{1}>$,
 $T_{2}$ as $_{2}<\vec{x}\vec{y}|T_{2}>$ and
$T_{3}$ as $_{3}<\vec{x}\vec{y}|T_{3}>$. Then after inserting completeness
relations the first term on the right hand side  of Eq.(\ref{17}) becomes
\begin{eqnarray}
_{1}<\vec{x}\ \vec{y}\vert t_ 1 G_ 0 \vert T_ 2 + T_ 3 >  &  = & \int d \vec x' d \vec y'
 <\vec{x}\ \vec{y} \vert t_ 1 \vert \vec{x}^{\prime}\vec{y}^{\prime}>_{1}
\int d \vec x'' d \vec y'' {_{1}<\vec{x}^{\prime}\ \vec
{y}^{\prime}\vert G_{0}\vert \vec{x}^{\prime\prime}\vec{y}^{\prime\prime}>_{1}} \label{52}\\
&  & \int d \vec x''' d \vec y''' [_{1}<\vec{x}^{\prime\prime\ }\vec{y}^{\prime\prime}|\vec{x}^{\prime
\prime\prime}\ \vec{y}^{\prime\prime\prime}>_{2}{_{2}<\vec{x}^{\prime
\prime\prime}\ \vec{y}^{\prime\prime\prime}|T_{2}>}\nonumber\\
&  + & _{1}<\vec{x}^{\prime\prime}\ \vec{y}^{\prime\prime}|\vec{x}^{\prime
\prime\prime}\ \vec{y}^{\prime\prime\prime}>_{3}{_{3}<\vec{x}^{\prime
\prime\prime}\ \vec{y}^{\prime\prime\prime}|T_{3}>}]\nonumber
\end{eqnarray}
Various matrix elements occur which we now consider one by one. The 2-body
t-matrix obeys the Lippmann Schwinger equation (\ref{13}), which in configuration
space reads
\begin{eqnarray}
<\vec{x}\ \vec{y}|t|\vec{x}^{\prime}\ \vec{y}^{\prime}>  &  = & <\vec{x}\ \vec
{y}|V|\vec{x}^{\prime}\ \vec{y}^{\prime}> \label{53}\\
&  + & \int d \vec x'' d \vec y'' 
<\vec{x}\ \vec{y}|V|\vec{x}^{\prime\prime}\ \vec{y}^{\prime\prime }>
\int d \vec x''' d \vec y''' <\vec{x}^{\prime\prime}\ \vec{y}^{\prime\prime}|G_{0}|\vec{x}^{\prime
\prime\prime}\ \vec{y}^{\prime\prime\prime}><\vec{x}^{\prime\prime\prime
}\ \vec{y}^{\prime\prime\prime}|t|\vec{x}^{\prime}\ \vec{y}^{\prime}>\nonumber
\end{eqnarray}

The two-atom force $V$ is diagonal in the vector $\vec{y}$ which means
\begin{eqnarray}
<\vec x  \vec y \vert V \vert \vec x'  \vec y' > = \delta(\vec y -\vec y') 
< \vec x \vert V \vert \vec x' > = \frac{1}{(2\pi)^{3}} \int d \vec q 
 e^{i \vec q \cdot( \vec y - \vec y')} < \vec x \vert V \vert \vec x'>  \label{54}
\end{eqnarray}
The three-body free  Green's function can be related to a two-body free Green's function,
 by making use of the property that each is diagonal in momentum space.  We use
Jacobi momenta $\vec{p}^{(i)},\vec{q}^{(i)}$ related to $\vec{x}^{(i)},\vec
{y}^{(i)}$ and obtain
\begin{eqnarray}
<\vec{x}\ \vec{y}|G_{0}|\vec{x}^{\prime}\ \vec{y}^{\prime}>  &  = & \int d\vec
{p}\ d\vec{q}<\vec{x}\ \vec{y}|\vec{p}\ \vec{q}>\frac{1}{E+i\epsilon
-\frac{p^{2}}{2\mu_{i}}-\frac{q^{2}}{2M_{i}}}<\vec{p}\ \vec{q}|\vec{x}
^{\prime}\ \vec{y}^{\prime}>\nonumber\\
&  =& \frac{1}{(2\pi)^{6}}\int d\vec{p}\ d\vec{q}\ e^{i\vec{p}\cdot(\vec{x}
-\vec{x}^{\prime})}e^{i\vec{q}\cdot(\vec{y}-\vec{y}^{\prime})}\frac
{1}{E+i\epsilon-\frac{p^{2}}{2\mu_{i}}-\frac{q^{2}}{2M_{i}}}\nonumber\\
&  \equiv & \frac{1}{(2\pi)^{3}}\int d\vec{q}\ e^{i\vec{q}\cdot(\vec{y}-\vec
{y}^{\prime})}g(\vec{x},\vec{x}^{\prime};\epsilon_{q}) \label{55}
\end{eqnarray}
where $g(\vec{x},\vec{x}^{\prime};\epsilon_{q})$ is the well known free single
particle Greens function
\begin{eqnarray}
g(\vec{x},\vec{x}^{\prime};\epsilon_{q})=-\frac{\mu_i}{2 \pi}
\frac{e^{i\sqrt{2\mu_{i}(E - \frac{q^2}{2M_{i}})}}|\vec{x}-\vec{x}^{\prime}|}{|\vec
{x}-\vec{x}^{\prime}|}
\end{eqnarray}
We introduced the reduced mass $\mu_{i}$ of the two atoms in the arrangement (i)
\begin{eqnarray}
\mu_i = \frac{m_j m_k}{m_j + m_k}
\end{eqnarray}
and $ M_i$ the reduced mass between the particle i and the pair ( jk)
\begin{eqnarray}
M_i = \frac{m_i ( m_j+m_k)}{m_i + m_j + m_k}
\end{eqnarray} 
and  $\epsilon_{q}\equiv
E-\frac{q^{2}}{2M_{i}}$ is the energy related to the 2- atom subsystem.

Similarily, we can relate the two-body t-matrix embedded in the three-body space 
to the two-body $ \tau$-matrix defined in the two-body space as follows.
Inserting (\ref{54}) and (\ref{55}) into the LSE (\ref{53}) 
it follows that the t-matrix  has
the form
\begin{eqnarray}
<\vec{x}\vec{y}|t|\vec{x}^{\prime}\vec{y}^{\prime}>\equiv\frac{1}{(2\pi)^{3}
}\int d\vec{q}e^{i\vec{q}\cdot(\vec{y}-\vec{y}^{\prime})}<\vec{x}
|\tau(\epsilon_{q})|x^{\prime}> \label{59}
\end{eqnarray}
where $\tau(\vec{x},\vec{x}^{\prime};\epsilon_{q})\equiv<\vec{x}|\tau
(\epsilon_{q})|x^{\prime}>$ obeys the two-body  LSE
\begin{eqnarray}
\tau(\vec{x},\vec{x}^{\prime};\epsilon_{q})=V(x)\delta(\vec{x}-\vec{x}%
^{\prime})+\int d\vec{x}^{\prime\prime}V(x)g(\vec{x},\vec{x}^{\prime\prime
};\epsilon_{q})\tau(\vec{x}^{\prime\prime},\vec{x}^{\prime};\epsilon_{q})
\label{60},
\end{eqnarray}
and where we have assumed that $ V(x)$ is  a local potential.

The recoupling matrix elements in (\ref{52}) requires some consideration. The three
different sets of Jacobi vectors are linearily related to each other. Thus
$\vec{x}^{(2)}$ and $\vec{y}^{(2)}$ in arrangement 2
 can be expressed in terms of $\vec
{x}^{(1)}$ and $\vec{y}^{(1)}$ in arrangement 1 as
\begin{align}
\vec{x}^{(2)}  &  =A\vec{x}^{(1)}+B\vec{y}^{(1)}\\
\vec{y}^{(2)}  &  =C\vec{x}^{(1)}+D\vec{y}^{(1)}%
\end{align}
 Similarily
\begin{align}
\vec{x}^{(3)}  &  =A^{\prime}\vec{x}^{(1)}+B^{\prime}\vec{y}^{(1)}\\
\vec{y}^{(3)}  &  =C^{\prime}\vec{x}^{(1)}+D^{\prime}\vec{y}^{(1)}%
\end{align}
where 
\begin{eqnarray}
 A & = &  -\frac{m_2}{m_2 + m_3} ~~~~~~~ B= -1\\
 C & = &  \frac{m_3 }{ m_2 + m_3 } \frac{m_1 + m_2 + m_3 }{m_1 + m_3}~~~~~~D = - \frac{m_1 }{m_1 + m_3}\\
 A' & = &  -\frac{m_3 }{m_2 + m_3} ~~~~~~~~ B'= 1\\
 C' & = &  - \frac{m_2 }{m_2 + m_3} \frac{m_1 + m_2 + m_3 }{ m_1 + m_2} ~~~~~~D'= - \frac{m_1}{m_1 + m_2}
 \end{eqnarray}

As a consequence one can express the state $|\vec{x}\ \vec{y}>_{1}$ as
\begin{align}
|\vec{x}\ \vec{y}  &  >_{1}=|A\vec{x}+B\vec{y},C\vec{x}+D\vec{y}>_{2}\\
|\vec{x}\ \vec{y}  &  >_{1}=|A^{\prime}\vec{x}+B^{\prime}\vec{y},C^{\prime
}\vec{x}+D^{\prime}\vec{y}>_{3}%
\end{align}
The above equations mean that a spatial configuration represented in arrangement 1 by
vectors $ \vec x $ and $ \vec y$ is represented in arrangement 2 by vectors 
$ A \vec x + B \vec y$ and $ C \vec x + D \vec y$ and similarily for arrangement 3.  
This leads to
\begin{eqnarray}
_{1}  &  <\vec{x}\ \vec{y}|\vec{x}^{\prime}\ \vec{y}^{\prime}>_{2}=\delta
(\vec{x}^{\prime}-A\vec{x}-B\vec{y})\delta(\vec{y}^{\prime}-C\vec{x}-D\vec
{y}) \label{67}\\
_{1}  &  <\vec{x}\ \vec{y}|\vec{x}^{\prime\ }\vec{y}^{\prime}>_{3}=\delta
(\vec{x}^{\prime}-A^{\prime}\vec{x}-B^{\prime}\vec{y})\delta(\vec{y}^{\prime
}-C^{\prime}\vec{x}-D^{\prime}\vec{y}) \label{68}
\end{eqnarray}
Using now Eqs(\ref{52}),(\ref{59}), (\ref{55}), (\ref{67}), (\ref{68}) we obtain
\begin{eqnarray}
_{1}<\vec{x}\ \vec{y}|t_{1}& G_0 & \vert T_{2}+T_{3}>    = 
 \frac{1}{(2\pi)^{3}}\int d\vec{q} \int d\vec{y}^{\prime\prime} \int d\vec
{x}^{\prime\prime}\\
& & e^{i\vec{q}\cdot(\vec{y}-\vec y'')} \int d\vec{x}^{\prime}\tau(\vec{x}
,\vec{x}^{\prime};\epsilon_{q})g(\vec{x}^{\prime},\vec{x}^{\prime\prime
};\epsilon_{q})\nonumber\\
&   & [ _2<A\vec{x}^{\prime\prime}+B\vec{y}^{\prime\prime},C\vec{x}^{\prime\prime
}+D\vec{y}^{\prime\prime}|T_{2}>
  +  _3<A^{\prime}\vec{x}^{\prime\prime}+B^{\prime}\vec{y}^{\prime\prime
},C^{\prime}\vec{x}^{\prime\prime}+D^{\prime}\vec{y}^{\prime\prime}|T_{3}>]\nonumber
\end{eqnarray}
Thus the unknown amplitudes $_{2}<\vec{x}\ \vec{y}|T_{2}>$ and $_{3}<\vec
{x}\ \vec{y}|T_{3}>$ occur with shifted arguments under the integral.

Next we regard the second term on the right hand side of Eq.(\ref{17}):
\begin{eqnarray}
_1<\vec{x}\ \vec{y}|(1+& t_{1} & G_{0})V_4^{(1)}|\phi_{1}>    = \int d \vec x' d \vec y' 
 _1<\vec{x}\ \vec
{y}|V_4^{(1)}|\vec{x}^{\prime}\vec{y}^{\prime}>_{1}{_{1}<\vec{x}^{\prime}
\ \vec{y}^{\prime}|\phi_{1}>}\\
&  + &\int d \vec x' d \vec y'
 _1<\vec{x}\ \vec{y}|t_{1}G_{0})|\vec{x}^{\prime}\ \vec{y}^{\prime}>_1 
\int d \vec x'' d \vec y'' {_1<\vec{x}^{\prime}\ \vec{y}^{\prime}|V_4^{(1)}|\vec{x}^{\prime\prime}\ \vec{y}
^{\prime\prime}>_1} {_1<\vec{x}^{\prime\prime}\ \vec{y}^{\prime\prime}|\phi_{1}>_1}\nonumber
\end{eqnarray}

The 3- atom force $V_4^{(1)}$ has to be a scalar under rotations. Therefore for
spinless atoms and assuming locality it will have the form
\begin{eqnarray}
_{1}<\vec{x}\ \vec{y}|V_4^{(1)}|\vec{x}^{\prime}\ \vec{y}^{\prime}>_{1}
=V^{(1)}(x,y,\hat{x}\cdot\hat{y})\delta(\vec{x}-\vec{x}^{\prime})\delta
(\vec{y}-\vec{y}^{\prime}) \label{71}
\end{eqnarray}
Therefore,  together with (\ref{71}),(\ref{59}),(\ref{55}) we obtain
\begin{eqnarray}
<\vec{x}\ \vec{y}|(1+t_{1}G_{0})V_4^{(1)}|\phi_{1}>  &  =V^{(1)}(x,y,\hat
{x}\cdot\hat{y})u(\vec{x})\frac{1}{(2\pi)^{3/2}}e^{i\vec{q}_{0}\cdot\vec{y}%
}\nonumber\\
&  +\frac{1}{(2\pi)^{3}}\int d\vec{q} \int d\vec{x}^{\prime}
\int d\vec{y}^{\prime}e^{i\vec{q}\cdot(\vec{y}-\vec{y}^{\prime})}
\int d\vec{x}^{\prime\prime} \tau(\vec{x},\vec{x}^{\prime\prime};\epsilon_{q})\nonumber\\
&  \times g(\vec{x}^{\prime\prime},\vec{x}^{\prime};\epsilon_{q}%
)V^{(1)}(x^{\prime},y^{\prime},\hat{x}^{\prime}\cdot\hat{y}^{\prime})u(\vec
{x}^{\prime})\frac{1}{(2\pi)^{3/2}}e^{i\vec{q}_{0}\cdot\vec{y}^{\prime}}
\label{72}
\end{eqnarray}

Finally the third term on the right hand side of Eq. (\ref{17}) can be transformed by means of
similar steps with the result
\begin{eqnarray}
< \vec{x}\ \vec{y}|(1+t_{1}G_{0}) &  & V_4^{(1)}G_{0}(T_{1}+T_{2}+T_{3})>\\
 & = & V^{(1)}( x,y,\hat x \cdot \hat y) \frac{1}{(2 \pi)^3} \int d \vec q
\int d \vec x' \int d \vec y'  e^{ i \vec q \cdot ( \vec y - \vec y')} g( \vec x \vec x'; \epsilon_q)\nonumber\\
& &  ( < \vec x' \vec y' \vert T_1> +
 < A \vec x' + B \vec y', C \vec x' + D \vec y' \vert T_2>\nonumber\\ 
& + & < A' \vec x' + B' \vec y', C' \vec x' + D' \vec y' \vert T_3>) \nonumber\\
& + & \frac{1}{(2 \pi)^3} \int d \vec q \int d \vec y''
 e^{ i \vec q  \cdot ( \vec y - \vec y'')} \int d \vec x'' \int d \vec x'
\tau ( \vec x, \vec x';\epsilon_q)  g( \vec x', \vec x''; \epsilon_q)\nonumber\\ 
& & V^{(1)}( x'',y'',\hat x'' \cdot \hat y'') \frac{1}{(2 \pi)^3}
 \int d \vec q' \int d \vec x''' \int d \vec y''' 
 e^{ i \vec q'  \cdot ( \vec y'' - \vec y''')}
 g( \vec x'' \vec x'''; \epsilon_{q'})\nonumber\\
& &  ( < \vec x''' \vec y''' \vert T_1> +
 < A \vec x''' + B \vec y''', C \vec x''' + D \vec y''' \vert T_2>\nonumber\\
&  + & < A' \vec x''' + B' \vec y''', C' \vec x''' + D' \vec y''' \vert T_3>) \nonumber 
\end{eqnarray}

We leave it to the reader to work out the coordinate space representations for
the remaining two equations (\ref{18}), ( \ref{19}).

In case of identical particles there is only one FE given by  (\ref{41})  to be solved for the
amplitude $T( \vec x, \vec y) \equiv< \vec x \vec y \vert T> $. This eqation is of the form
\begin{eqnarray}
 T = T^0 + K \label{int}
\end{eqnarray}
 where $ T^0$ is the driving term and $ K$ incorporates the  $T$- amplitudes.
In the coordinate space representation
\begin{eqnarray}
T^0 ( \vec x, \vec y) &  = &  (\frac{4}{3})^3 (\frac{1}{2 \pi})^{9/2}
 \int d \vec y'  \int d \vec y'' \int d \vec q
 e^{ i \vec q  \cdot ( \vec y - \vec y')} \label{74a}\\
& & [ \tau ( \vec x, \frac{4}{3} \vec y'' + \frac{2}{3} \vec y'; \epsilon_q)
u_{n_0} ( -\frac{2}{3} \vec y'' - \frac{4}{3} \vec y')
  +   \tau ( \vec x,- \frac{4}{3} \vec y'' - \frac{2}{3} \vec y'; \epsilon_q)
u_{n_0} ( \frac{2}{3} \vec y'' + \frac{4}{3} \vec y')] 
 e^{ i \vec q_0  \cdot  \vec y''}\nonumber\\
&  + & V^{(1)} ( x,y,\hat x \cdot \hat y) \frac{1}{(2 \pi)^{3/2}} [ u_{n_0}( \vec x)
 e^{i \vec q_0 \cdot \vec y}\nonumber\\
&  + &  u_{n_0} ( -\frac{1}{2} \vec x -  \vec y) 
e^{i \vec q_0 \cdot ( \frac{3}{4} \vec x - \frac{1}{2} \vec y)}
 +   u_{n_0} ( -\frac{1}{2} \vec x +  \vec y)
e^{i \vec q_0 \cdot ( -\frac{3}{4} \vec x - \frac{1}{2} \vec y)}]\nonumber\\
& + &  \frac{1}{(2 \pi)^3} \int d \vec q  \int d \vec y'' \int d \vec x'' 
e^{i \vec q \cdot( \vec y - \vec y'')} 
 \int d \vec x'\tau( \vec x, \vec x'; \epsilon_q) g( \vec x', \vec x''; \epsilon_q)\nonumber\\
& &  V^{(1)} ( x'',y'',\hat x'' \cdot \hat y'') \frac{1}{( 2 \pi)^{3/2}} 
[ ( u_{n_0}( \vec x'')  e^{i \vec q_0 \cdot \vec y''}\nonumber\\
&  + &  u_{n_0} ( -\frac{1}{2} \vec x'' -  \vec y'')
e^{i \vec q_0 \cdot ( \frac{3}{4} \vec x'' - \frac{1}{2} \vec y'')}
 +   u_{n_0} ( -\frac{1}{2} \vec x'' +  \vec y'') e^{i \vec q_0 \cdot ( -\frac{3}{4} \vec x''
 - \frac{1}{2} \vec y'')}]\nonumber
\end{eqnarray}
and 
\begin{eqnarray}
K( \vec x \vec y) & =  &   \frac{1}{(2 \pi)^3}  \int d \vec x'' \int d \vec y'' \int d \vec q 
 e^{ i \vec q  \cdot ( \vec y - \vec y'')}
  \int d \vec x'  \tau ( \vec x,  \vec x';  \epsilon_q)
 g( \vec x',  \vec x'';  \epsilon_q)\label{74b}\\
& & [ T( -\frac{1}{2} \vec x'' - \vec y'',
 \frac{3}{4} \vec x'' - \frac{1}{2} \vec y'') +
 T( -\frac{1}{2} \vec x'' + \vec y'',
 -\frac{3}{4} \vec x'' - \frac{1}{2} \vec y'')] \nonumber\\
& + & V^{(1)} ( x,y,\hat x \cdot \hat y)  \frac{1}{(2 \pi)^3}  \int d \vec q
\int d \vec y'  e^{ i \vec q  \cdot  ( \vec y - \vec y')}
 \int d \vec x'  g( \vec x, \vec x'; \epsilon_q) 
[ < \vec x' \vec y' \vert T>\nonumber\\
&  + &  < A \vec x' + B \vec y', C \vec x' + D \vec y' \vert T>
 +   < A' \vec x' + B' \vec y', C' \vec x' + D' \vec y' \vert T>]\nonumber\\
& + &   \frac{1}{(2 \pi)^3} \int d \vec q \int d \vec y'' \int d \vec x'
e^{ i \vec q  \cdot  ( \vec y - \vec y')}
 \int d \vec x''\tau(\vec x, \vec x''; \epsilon_q) g( \vec x'', \vec x'; \epsilon_q)
  V^{(1)} ( x',y',\hat x' \cdot \hat y')\nonumber\\
& &   \frac{1}{(2 \pi)^3} \int d \vec q' 
e^{ i \vec q'  \cdot  ( \vec y' - \vec y'')} g( \vec x', \vec x''; \epsilon_{q'})\nonumber\\
& & [  < \vec x'' \vec y'' \vert T> + < A \vec x'' + B \vec y'', C \vec x'' + D \vec y'' \vert T>\nonumber\\
& + &  < A' \vec x'' + B' \vec y'', C' \vec x'' + D' \vec y'' \vert T>]\nonumber 
\end{eqnarray} 

Once the amplitude $T(\vec{x},\vec{y})$ has been determined the physical $3-$
atom break up matrix element is obtained as
\begin{align}
<\vec{p}\ \vec{q}|U_{0}>  &  \equiv\ <\vec{p}\ \vec{q}|(1+\mathcal{P})T>\\
&  =\frac{1}{(2\pi)^{3}}\int d\vec{x}\ d\vec{y}\ e^{-i\vec{p}\cdot\vec{x}%
}e^{-i\vec{q}\cdot\vec{y}\ }T(\vec{x},\vec{y})+permuted~~parts
\end{align}
The permuted parts are most conveniently evaluated by applying the permutation
$\mathcal{P}$  to the left:
\begin{eqnarray}
_{1}<\vec{p}\ \vec{q}\vert \mathcal{P}=_{2}<\vec{p}\ \vec{q}|+_{3}<\vec{p}\ \vec{q}\vert
\end{eqnarray}
Then one expresses the Jacobi momenta of the type 1 in terms of the Jacobi
momenta of the types 2 and 3. One has in case of identical particles\cite{gloeckle83}
\begin{eqnarray}
\vec p^{(1)}    = -\frac{1}{2} \vec p^{(2)} + \frac{3}{4} \vec q^{(2)}\\
\vec q^{(1)}    = - \vec p^{(2)} -\frac{1}{2} \vec q^{(2)}\\
\vec p^{(1)}    = -\frac{1}{2} \vec p^{(3)} -\frac{3}{4} \vec q^{(3)}\\
\vec q^{(1)}    =  \vec p^{(3)} -\frac{1}{2} \vec q^{(3)}
\end{eqnarray}

and therefore
\begin{eqnarray}
_{2}< \vec p   \vec q \vert =_{1}< -\frac{1}{2} \vec p + \frac{3}{4} \vec q, 
 - \vec p -\frac{1}{2} \vec q \vert \label{82}
\end{eqnarray}

and similarily for $_{3}<\vec{p}\vec{q}\vert$ .
 Thus altogether
\begin{eqnarray}
<\vec{p}\ \vec{q}|U_{0}>=T(\vec{p}\ \vec{q}) + T( -\frac{1}{2} \vec p + \frac{3}{4} \vec q,
- \vec p -\frac{1}{2} \vec q)  + T(-\frac{1}{2} \vec p -\frac{3}{4} \vec q,
\vec p -\frac{1}{2} \vec q)
\end{eqnarray}

It remains to display the physical matrix element $< \phi_{1}^{\prime}\vert U>
$ for elastic and inelastic  scattering , where $ \vert \phi_1' > = \vert u'> \vert \vec q'> $
 represents the final state.
 According to (\ref{43}) one obtains
\begin{eqnarray}
< \phi_{1}^{\prime}\vert U>  &  =  & < \phi_{1}^{\prime}\vert \mathcal{P}G_{0}^{-1}
\vert\phi_{1}> + < \phi_{1}^{\prime}\vert \mathcal{P} \vert T > \label{84}\\
&  + & < \phi_{1}^{\prime}\vert V_4^{(1)} ( 1+\mathcal{P} ) \vert\phi_{1}> + < \phi
_{1}^{\prime}\vert V^{(1)} ( 1+\mathcal{P}) G_{0} T>\nonumber
\end{eqnarray}

For the first term we can use the SE for $\phi_{1} $ and obtain
\begin{eqnarray}
< \phi_{1}^{\prime}\vert \mathcal{P}G_{0}^{-1} \vert\phi_{1}>  &  = & < \phi_{1}^{\prime
}\vert \mathcal{P} V \vert\phi_{1}>\\
&  =  & \int d \vec x^{\prime}d \vec y^{\prime}d \vec x^{\prime\prime}d \vec
y^{\prime\prime}< \phi_{1}^{\prime}\vert\vec x^{\prime}\vec y^{\prime}>_{1}
{_{1}< \vec x^{\prime}\vec y^{\prime}\vert \mathcal{P} \vert\vec x^{\prime\prime}\vec
y^{\prime\prime}>_{1}} { _{1} < \vec x^{\prime\prime}\vec y^{\prime\prime
}\vert V \vert\phi_{1}>}\nonumber
\end{eqnarray}

This is an example where one might evaluate the $\mathcal{P}$ -matrix element differently
from the one shown above. The expressions (\ref{67}),(\ref{68}) can be rewritten as
\begin{eqnarray}
_{1}<\vec{x}\ \vec{y}|\mathcal{P}|\vec{x}^{\prime}\ \vec{y}^{\prime}>_{1} & = & 
(\frac{1}{C})^{3} \delta( \vec x - \frac{1}{C} \vec y' - \frac{D}{C} \vec y)
\delta( \vec x' - \frac{A}{C} \vec y' -  \frac{A D- BC}{C} \vec y)\\
 & + & (\frac{1}{C'})^{3} \delta( \vec x - \frac{1}{C'} \vec y' - \frac{D'}{C'} \vec y)
\delta( \vec x' - \frac{A'}{C'} \vec y' -  \frac{A' D'- B'C'}{C'} \vec y)\nonumber
\end{eqnarray}

 For identical atoms one has $ A=-\frac{1}{2}, B=-1,C=\frac{3}{4}, D= -\frac{1}{2},
A'= -\frac{1}{2}, B'= 1, C'= -\frac{3}{4}, D'= -\frac{1}{2}$.
n
Then we obtain
\begin{eqnarray}
< \phi_{1}^{\prime} \vert \mathcal{P} G_{0}^{-1} \vert \phi_{1}> & = &
 < \phi_{1}^{\prime} \vert \mathcal{P} V \vert \phi_{1}>\\
& = & \frac{1}{(2 \pi)^3} ( \frac{4}{3})^3 \int d \vec y \int d \vec y'
 e^{- i \vec q' \cdot \vec y } e^{ i \vec q_0 \cdot \vec y' }\\
& & ( u_{n'}^{*} ( \frac{4}{3} \vec y' + \frac{2}{3} \vec y)
  V( \vert \frac{2}{3} \vec y' + \frac{4}{3} \vec y \vert ) 
u_{n_0} ( - \frac{2}{3} \vec y' - \frac{4}{3} \vec y)\nonumber\\
& + &  u_{n'}^{*} (- \frac{4}{3} \vec y' - \frac{2}{3} \vec y)
  V( \vert \frac{2}{3} \vec y' + \frac{4}{3} \vec y \vert )
u_{n_0} (  \frac{2}{3} \vec y' + \frac{4}{3} \vec y))\nonumber
\end{eqnarray}

This together with the remaining terms on the right hand side of (\ref{84})  yields
\begin{eqnarray}
 & < & \phi_{1}^{\prime}|U>   =   \frac{1}{(2 \pi)^3} ( \frac{4}{3})^3
 \int d \vec y \int d \vec y'  e^{- i \vec q' \cdot \vec y } e^{ i \vec q_0 \cdot \vec y' }\\
& & [ u_{n'}^{*} ( \frac{4}{3} \vec y' + \frac{2}{3} \vec y)
  V( \vert \frac{2}{3} \vec y' + \frac{4}{3} \vec y \vert )
u_{n_0} ( - \frac{2}{3} \vec y' - \frac{4}{3} \vec y)
 +   u_{n'}^{*} (- \frac{4}{3} \vec y' - \frac{2}{3} \vec y)
  V( \vert \frac{2}{3} \vec y' + \frac{4}{3} \vec y \vert )
u_{n_0} (  \frac{2}{3} \vec y' + \frac{4}{3} \vec y)]\nonumber\\
& + &  \frac{1}{(2 \pi)^{3/2}} ( \frac{4}{3 })^3 \int d \vec y \int d \vec y' 
 e^{- i \vec q' \cdot \vec y }  
  [ u_{n'}^{*} ( \frac{4}{3} \vec y' + \frac{2}{3} \vec y)
  T( - \frac{2}{3} \vec y' - \frac{4}{3} \vec y, \vec y' )
 +   u_{n'}^{*} (- \frac{4}{3} \vec y' - \frac{2}{3} \vec y)
  T(  \frac{2}{3} \vec y' + \frac{4}{3} \vec y, \vec y' ]\nonumber\\
&  + &  \frac{1}{(2 \pi)^3} e^{ -i \vec q' \cdot \vec y} u_{n'}^{*} ( \vec x )
 V^{(1)} ( x y , \hat x \cdot \hat y)   [ u_{n_0} ( \vec x ) e^{i \vec q_0 \cdot \vec y}\nonumber\\
&  + & u_{n_0} ( -\frac{1}{2}\vec x - \vec y) 
e^{i \vec q_0 \cdot ( \frac{3}{4} \vec x - \frac{1}{2} \vec y)}
 +   u_{n_0} ( -\frac{1}{2}\vec x + \vec y)
e^{i \vec q_0 \cdot ( -\frac{3}{4} \vec x - \frac{1}{2} \vec y)}]\nonumber\\
& + &  \frac{1}{( 2 \pi)^{9/2}} e^{ -i \vec q' \cdot \vec y}u_{n'}^{*} ( \vec x)
 V^{(1)} ( x y , \hat x \cdot \hat y) \int d \vec q \int d \vec y' \int d \vec x' 
e^{ i \vec q \cdot (  \vec y - \vec y') }
g(\vec x, \vec x'; \epsilon_q)\nonumber\\
& &  ( < \vec x' \vec y' \vert T>
+ < A \vec x' + B \vec y' , C \vec x' + D \vec y' \vert T>
+ < A' \vec x' + B' \vec y' , C' \vec x' + D' \vec y' \vert T>)\nonumber
\end{eqnarray}

For distinguishable particles corresponding results occur and their evaluation
is left to the reader.\bigskip

\section{Partial wave decomposition}

In view of the various vibration-rotational levels it is natural to decompose
the integral equations for the various T-amplitudes into partial waves. We
exemplifiy that step for the case of three indentical atoms, and choose
therefore Eqs.(\ref{int}),(\ref{74a}),(\ref{74b}) .
 The  scalar quantities  $\tau$ as well as the single particle Green's
function $g$, both contained in Eqs. (\ref{int}),(\ref{74a}),(\ref{74b})
 have the partial wave
decomposition
\begin{eqnarray}
\tau(\vec{x},\vec{x}^{\prime};\varepsilon_{q})=\sum_{l,m}\ Y_{l,m}(\hat
{x})\tau_{l}(x,x^{\prime};\varepsilon_{q})\ Y_{l,m}^{\ast}(\hat{x}^{\prime})
\label{tau_l}
\end{eqnarray}
\begin{eqnarray}
g(\vec{x},\vec{x}^{\prime};\epsilon_{q})=\sum_{l,m}\ Y_{l,m}(\hat{x}
)g_{l}(x,x^{\prime};\varepsilon_{q})\ Y_{l,m}^{\ast}(\hat{x}^{\prime})
\label{g_l}
\end{eqnarray}
and
\begin{eqnarray}
\delta(\vec{x}-\vec{x}^{\prime})=\frac{\delta(x-x^{\prime})}{xx^{\prime}}
\sum_{l,m}\ Y_{l,m}(\hat{x})\ Y_{l,m}^{\ast}(\hat{x}^{\prime}) \label{delta_l}
\end{eqnarray}
 Here $\hat v $ is the unit vector pointing into the direction
of $ \vec v$. When used as an argument of a  spherical harmonics $ \hat v$ stands for
the angles $ \theta $ and $ \phi$.

As a consequence the LSE for $\tau$, Eq.(\ref{60}), for a given angular
momentum $l$ reads
\begin{eqnarray}
\tau_{l}(x,x^{\prime};\varepsilon_{q})  &   =& V\left(  x\right)  \frac
{\delta(x-x^{\prime})}{xx^{\prime}}\nonumber\\
&  + & V(x)\int_0^{\infty} dx''  x''^2 g_{l}(x,x^{\prime\prime
};\varepsilon_{q})\ \tau_{l}(x^{\prime\prime},x^{\prime};\varepsilon_{q})
\label{TAU_L_LSE}
\end{eqnarray}
Further ingredients are the initial bound state
\begin{eqnarray}
u_{n_{0}}(\vec{x})=u_{n_{0}}(x)\ Y_{l_{0},m_{0}}(\hat{x}) \label{BS0}
\end{eqnarray}
and the standard expansion of the plane wave
\begin{eqnarray}
e^{i\ \vec{q}\cdot\vec{y}}=4\pi\sum_{\lambda,\mu}\ i^{\lambda}j_{\lambda
}(qy)Y_{\lambda,\mu}^{\ast}(\hat{q})\ Y_{\lambda,\mu}(\hat{y}). \label{PW}
\end{eqnarray}
Next one introduces bi-polar spherical harmonics of total angular momentum
$L$
\begin{eqnarray}
\mathcal{Y}_{l,\lambda}^{\ L,M}(\hat{x},\hat{y})\equiv\sum_{m,\mu}C(l\lambda
L,m\mu M)Y_{l,m}(\hat{x})\ Y_{\lambda,\mu}(\hat{y}). \label{BPSH}
\end{eqnarray}
 where the $ C(\cdots)$  is a Clebsch- Gordan coefficients as defined in \cite{rose}. However,
in what follows we are going to omit the third magnetic quantum number, since $ m+ \mu = M$.

Using the notation above we will now begin to evaluate the amplitude T given in 
Eqs. (\ref{41}) and (\ref{int}),(\ref{74a}),(\ref{74b}).
The two driving terms in the first line of Eq(\ref{41})
 will be denoted by $ T^{0,1}$ and $ T^{0,2}$, respectively.  

As an intermediate result one obtains for $ T^{0,1}$ 
\begin{eqnarray}
 < \vec x \vec y \vert T^{0,1} >  & \equiv &   < \vec x \ \vec y \vert t\mathcal{P}| \phi>\\
 &  = & (\frac{4}{3})^{3}\frac{1}{\left(  2\pi\right)
^{9/2}}(4\pi)^{3} \sum_{l \lambda} \sum_{LM} \ \mathcal{Y}_{l,\lambda}^{L,M}
(\hat x ,\hat y )\nonumber\\
& &  \times\int_{0}^{\infty}\ dq  q^2 j_{\lambda}(qy) \ \int d\ \vec y' 
\ \int d \vec y'' j_{\lambda}(qy')
 \mathcal{Y}_{l,\lambda}^{\ast L,M}
(\widehat{ \frac{4}{3}\vec y'' + \frac{2}{3}\vec y'}, \hat y')\nonumber\\
&  & \times\tau_{l}(x,\vert \frac{4}{3}\vec y'' + \frac{2}{3} \vec y' \vert ;
 \varepsilon_{q})
\ u_{n_{0}}(\vert  \frac{2}{3} \vec y'' + \frac{4}{3} \vec y' \vert)
\ \sum_{\lambda_0 \mu_ 0 } \ i^{\lambda_0 } j_{\lambda_{0}}
(q_{0}y'')Y_{\lambda_{0},\mu_{0}}^{\ast}(\hat{q}_{0})\nonumber\\
&  & \sum_{L'} \ C(l_0 \lambda_0 L', m_0, \mu_0)
\mathcal{Y}_{l_0,\lambda_0}^{ L', m_0 + \mu_0}
 (\widehat{ \frac{2}{3} \vec y'' + \frac{4}{3} \vec y'}, \hat y'') 
\ [(-)^{l_0}+(-)^{l}], \label{97}
\end{eqnarray}
n
Next we expand $\tau_{l}(x,\vert \frac{4}{3}\vec y'' + \frac{2}{3} \vec y' \vert;
\varepsilon_{q})
 u_{n_{0}}( \vert \frac{2}{3} \vec y'' + \frac{4}{3} \vec y' \vert )$
 into Legendre polynomials, using 
$P_{k}(\hat{x}\cdot \hat{y})= 
 (-)^k \frac{4 \pi}{\sqrt{\hat k}} \mathcal{Y}_{k,k}^{0,0}(\hat{x},\hat{y})$ and obtain
\begin{eqnarray}
\tau_{l}(x, \vert \frac{4}{3}\vec y'' + \frac{2}{3} \vec y' \vert ;
\varepsilon_{q})
 u_{n_{0}}(\vert  \frac{2}{3} \vec y'' + \frac{4}{3} \vec y' \vert )
=\sum
_{k}\ 2\pi\sqrt{2k+1}(-)^{k}\mathcal{Y}_{k,k}^{0,0}(\hat{y}^{\prime},\hat
{y}^{\prime\prime})\mathit{G}_{k}, \label{tau_u}
\end{eqnarray}
where
\begin{eqnarray}
\mathit{G}_{k}=\int_{-1}^{+1}dt\ P_{k}(t)\ \tau_{l}(x,\vert
\frac{4}{3}\vec y'' + \frac{2}{3} \vec y' \vert ;\varepsilon_{q}) 
 u_{n_{0}}(\vert \frac{2}{3} \vec y'' + \frac{4}{3} \vec y' \vert), \label{102}
\end{eqnarray}
and where $t$ is the cosine of the angle between the vectors $\vec y~'' $ and $\vec y~' $
 After
placing $\hat{q}_{0}$ into the $\hat{z}$ direction we  obtain
\begin{eqnarray}
  <\vec{x}\ \vec{y}|t\mathcal{P}|\phi> &  = & (\frac{4}{3})^{3}\frac{1}{\left(  2\pi\right)
^{9/2}}(4\pi)^{3}\sum_{l\lambda}\sum_{LM}\ \mathcal{Y}_{l,\lambda}^{L,M}
(\hat{x},\hat{y})\nonumber\\
& & \times\int_{0}^{\infty}\ dq\ q^{2}j_{\lambda}(qy)\ \int_{0}^{\infty}
d\ \vec{y}^{\prime}\ (y'^2 \int_{0}^{\infty}d\ \vec{y}^{\prime
\prime}y''^2 j_{\lambda}(qy^{\prime})\nonumber\\
& & \times\int d\hat{y}^{\prime}\ \int d\hat{y}^{\prime\prime}\mathcal{Y}
_{l,\lambda}^{\ast L,M}(\widehat{\frac{4}{3}\vec y'' + \frac{2}{3} \vec y' },
\hat{y}^{\prime})\ \sum_{k}\ (2\pi
)\sqrt{2k+1}(-)^{k}\mathcal{Y}_{k,k}^{0,0}(\hat{y}^{\prime},\hat{y}
^{\prime\prime})\mathit{G}_{k}\nonumber\\
&  & \times\ \sum_{\lambda_0} \ i^{\lambda_{0}}j_{\lambda_{0}}
(q_{0}y^{\prime\prime})\sqrt{(2\lambda_{0}+1)/(4\pi)}\nonumber\\
& &  \times\sum_{L^{\prime}}\ C(l_{0}\lambda_{0}L^{\prime},m_0,0)\ \mathcal{Y}
_{l_{0},\lambda_{0}}^{\ L',m_0} (\widehat{\frac{2}{3} \vec y'' + \frac{4}{3} \vec y'},
\hat{y}^{\prime\prime
})\ [(-)^{l_{0}}+(-)^{l}]. \label{xytP2}
\end{eqnarray}
 The remaining angular integrations can be performed analytically with the result
\begin{eqnarray}
  \int d \hat y'  \int d \hat y'' \mathcal{Y}_{l,\lambda}^{\ast
L,M}(\widehat{ \frac{4}{3}\vec y'' + \frac{2}{3} \vec y'},\hat{y}^{\prime})
& \mathcal{Y}_{k,k}^{0,0}& (\hat{y}^{\prime
},\hat{y}^{\prime\prime})  \mathcal{Y}_{l_{0},\lambda_{0}}^{\ L^{\prime}
,\mu_{0}}(\widehat{\frac{2}{3} \vec y'' + \frac{4}{3} \vec y'},\hat y'')\nonumber\\
&  = & \delta_{L,L^{\prime}}\ \delta_{M,m_0}\ h_{l\lambda,l_{0}\lambda_{0}
}^{k,L}(y^{\prime},y^{\prime\prime}) \label{104}
\end{eqnarray}
where the explicit expression for $ h $ is given in  Appendix A.

The $\tau$-  matrix occuring in Eq(\ref{102}) contains as a driving term the function
 $V\left(  x\right) \delta(x-x^{\prime})/x\ x^{\prime}$ ( see Eq.(\ref{TAU_L_LSE})).
 It is more convenient to explicitly separate this term out by defining the function
$r_{l}$
\begin{eqnarray}
\tau_{l}(x,x^{\prime};\varepsilon_{q})=V\left(  x\right)  \frac{\delta
(x-x^{\prime})}{xx^{\prime}}+r_{l}(x,x^{\prime};\varepsilon_{q}) \label{rl}
\end{eqnarray}
and treat the  $\delta$-function term $<\vec{x}\ \vec{y}|V\mathcal{P}|\phi>$ 
separately. Thus $ t\mathcal{P} \phi $
 will be decomposed into $  V\mathcal{P} \phi + r \mathcal{P} \phi$. One obtains
\begin{eqnarray}
  <\vec{x}\ \vec{y}|V\mathcal{P}|\phi> &  = & \sum_{l\lambda}\sum_{LM}\ \mathcal{Y}
_{l,\lambda}^{\ L,M}(\hat{x},\hat{y})\ \sqrt{2}\delta_{M,m_0}(-)^{l_{0}
}V(x)\nonumber\\
& &  \sum_{\lambda_{0}}\ i^{\lambda_{0}}\sqrt{2\lambda_{0}+1}C(l_{0}\lambda
_{0}L;m_0,0)\ \nonumber\\
& &  \sum_{k}\sqrt{2k+1}(-)^{k}\ \Lambda_{l\lambda,l_{0}\lambda_{0}}
^{k,L}(x,y)\ [1+(-)^{\lambda_{0}+\lambda+k}]\nonumber\\
& &  \int_{-1}^{+1}\ dt\ P_{k}(t)~u_{n_{0}}(\vert \frac{1}{2}\vec{x}+\vec{y} \vert)
 j_{\lambda_{0}}(q_{0}\vert \frac{3}{4}\vec{x}-\frac{1}{2}\vec{y} \vert), \label{xyVp}
\end{eqnarray}
 where $\Lambda$ is given by (see Appendix B)
\begin{eqnarray}
  \int d\hat{x}\ d\hat{y}\ \mathcal{Y}_{l,\lambda}^{\ast\ L,M}(\hat{x}
,\hat{y})\mathcal{Y}_{k,k}^{\ 0,0}(\hat{x},\hat{y})
\mathcal{Y}_{l_{0}\lambda_{0}}^{L^{\prime} m_0 }(\widehat{\frac{1}{2}\vec{x}+\vec{y}},
\widehat{\frac{3}{4}\vec{x}-\frac{1}{2}\vec{y}})\nonumber\\
  = \delta_{L,L^{\prime}}\delta_{M,m_0} \ \Lambda_{l\lambda,l_{0}
\lambda_{0}}^{k,L}(x,y). \label{109}
\end{eqnarray}
Now the projection onto the orthogonal states $\mathcal{Y}_{l,\lambda}
^{\ L,M}(\hat{x},\hat{y})$ defines $T_{l\lambda,LM}^{0,1}(x,y)$
\begin{eqnarray}
T_{l\lambda,LM}^{0,1}(x,y)\equiv\int d\hat{x}\ d\hat{y}\ \mathcal{Y}_{l,\lambda
}^{\ast\ L,M}(\hat{x},\hat{y})T^{0,1}(\vec{x},\vec{y}) \label{T0}
\end{eqnarray}
in terms of which
we obtain the two pieces of the driving term  $ T^{0,1}$
\begin{eqnarray}
T_{l \lambda LM}^{0,1}(xy) & = & \frac{2 \sqrt{2}}{\pi}(\frac{4}{3})^3((-)^{l_0}+(-)^l)
\delta_{M,m_0} \label{111}\\
& & \int_0^{\infty} dq  q^2 j_{\lambda}(qy)\int_0^{\infty} dy' y'^2
\int_0^{\infty} dy'' y''^2 j_{\lambda}(qy')\nonumber\\
& &  \sum_k \sqrt{\hat k}(-)^k \sum_{\lambda_0} i^{\lambda_0}
j_{\lambda_0}(q_0 y'')\sqrt{\hat \lambda_0}\nonumber\\
& &  h_{l \lambda,l_0 \lambda_0}^{ k,L}(y',y'')
C(l_0 \lambda_0 L,m_0 ,0)\nonumber\\
&  & \int_{-1}^{1} dt P_k(t) r_l(x,\vert \frac{4}{3} \vec y'' + \frac{2}{3}
\vec y' \vert;\epsilon_q ) u_{n_0}( \vert \frac{2}{3}\vec y'' + \frac{4}{3} \vec y' \vert)\nonumber\\
&  + & \sqrt{2} \delta_{M,m_0}(-)^{l_0} V(x)\sum_{\lambda_0}\sqrt {\hat \lambda_0}\nonumber\\
&   & C(l_0 \lambda_0 L,m_0, 0 ) \sum_{k}\sqrt{\hat k}(-)^k
\Lambda_{l \lambda, l_0 \lambda_0 }^{k,L}(x,y)\nonumber\\
&  & \int_{-1}^{1} dt P_k(t) u_{n_0}( \vert \frac{1}{2} \vec x+ \vec y \vert) j_{\lambda_0}
(q_0 \vert \frac{3}{4} \vec x - \frac{1}{2} \vec y \vert )(1+(-)^{\lambda_0 + \lambda + k})\nonumber
\end{eqnarray}
Here and in the following we use the notation $ \hat s \equiv 2 s +1 $.
 This symbol $ \hat s$ should of course not be confused with the unit vector.

n
 Before adressing the second part of the driving term in Eq(\ref{41}) or (\ref{74a})
 connected to the 3- atom force we regard the part of the kernel without 3-atom force,
 $ < \vec x \vec y \vert t G_0 \mathcal{P}T> $ . Using similar steps and projecting  onto states of
 total angular momentum L leads to
\begin{eqnarray}
 K_{l \lambda L}^{ (1)} ( xy) & \equiv &  \int d \hat x d \hat y \mathcal{Y}_{l \lambda}^{LM*}
 ( \hat x \hat y) < \vec x \vec y \vert t G_{0} \mathcal{P}T> \label{112}\\
  &  = &  4 \int_0^{\infty} dq q^{2} j_{\lambda} (qy)  \int_0^{\infty}
 dx'' x''^2  \int_0^{\infty} dx' x'^2
 \tau_{l}(x,x';\epsilon_{q}) g_{l}(x',x'';\epsilon_{q})\nonumber\\
& &  \int_0^{\infty} dy'' y'' j_{\lambda}(q y'') \sum_{l' \lambda'}  \sum_k
 \sqrt{\hat k}(-)^{l^{\prime}+k}
\Lambda_{l \lambda,l^{\prime}\lambda^{\prime}}^{k,L} ( x'' y'')\nonumber\\
&  & \int_{-1}^{1} dt P_{k}(t) T_{l^{\prime}\lambda^{\prime}~ L} ( \vert\frac{1}{2}
 \vec x''+ \vec y'' \vert, \vert \frac{3}{4}
 \vec x''- \frac{1}{2} \vec y'' \vert)
(1 + (-)^{\lambda^{\prime}+ \lambda})\nonumber
\end{eqnarray}

where we have defined the quantities $ T_{l \lambda L}$ in terms of the expansion
\begin{eqnarray}
T( \vec x \vec y) \equiv \sum_{l \lambda L M} {\cal Y}_{l \lambda}^{  LM}
 (\hat x  \hat y )T_{l \lambda L}(xy) \label{113}
\end{eqnarray}
 of $ T(\vec x \vec y) $ given in Eqs(\ref{41}) and (\ref{74b}).
We see that the unknown amplitudes $T_{l \lambda L}(x y)$ occur under the
integral with shifted arguments.

We are left with two more terms in Eqs.(\ref{41}), (\ref{74a}) and ( \ref{74b}),
 the second part of the 
 driving term and another
kernel piece, both containing the part $V_4^{(1)}$ of the 3-atom force. Let us
regard the simplest term $< \vec x \vec y \vert V^{(1)}\vert\phi_{1}>$ explicitely

We expand the plane wave and introduce states of total angular momentum, with the 
result
\begin{eqnarray}
< \vec x \vec y \vert V^{(1)}\vert\phi_{1}>  &  = &  \frac{4 \pi}{( 2\pi)^{3/2}}
\sum_{\lambda_{0}} i^{\lambda_{0}} \sqrt\frac{\hat\lambda_{0}}{4 \pi}
j_{\lambda_{0}} ( q_{0} y)   u_{n} ( x) V^{(1)} ( xy \hat x \cdot\hat y)\label{113a}\\
& & \sum_{L} C(l_0 \lambda_{0} L,m_0,0) \mathcal{Y}_{l_{0}
\lambda_{0}}^{L m_{0}} ( ,\hat x \hat y)\nonumber
\end{eqnarray}
Next we expand the 3-body force $V^{(1)} $ into Legendre polynomials
\begin{eqnarray}
V^{(1)}(x,y, \hat x \cdot \hat y) = \sum_{k} 2\pi\sqrt{\hat k} (-)^{k}
\mathcal{Y}_{kk}^{00}(\hat x \hat y) v_{k}
\end{eqnarray}
with
\begin{eqnarray}
v_{k}( x,y)  = \int_{-1}^{1} dt P_{k}(t) V^{(1)}(x,y, \hat x \cdot \hat y)
\end{eqnarray}

and combine the angular dependent terms as
\begin{eqnarray}
\mathcal{Y}_{kk}^{00}(\hat x \hat y) \mathcal{Y}_{l_{0} \lambda_{0}}^{L m_{0}
}(\hat x \hat y)  &  = & \frac{1}{4 \pi} \sqrt{ \hat k \hat l_{0} \hat
\lambda_{0}} (-)^{\lambda_{0} +L+k}\\
&  & \sum_{\mu_{1} \mu_{2}} (-)^{\mu_{1}}
\left\{ \begin{array}{ccc}
           \mu_1 & \mu_2 & L \\
           \lambda_0 & l_0 & k
          \end{array}
\right\} 
C(k l _{0} \mu_{1},00) C( k \lambda_{0} \mu_{2},00) \mathcal{Y}_{\mu_{1} \mu_{2}}^{L m_{0}}(\hat x \hat
y)\nonumber
\end{eqnarray}
 where the terms in curly bracket are 6-j symbols\cite{edmonds}.
Finally projecting Eq{\ref{113a})
 onto $\mathcal{Y}_{l \lambda}^{L M}(\hat x \hat y)$ one
obtains
\begin{eqnarray}
\int d \hat x & \int d \hat y & \mathcal{Y}_{l \lambda}^{L M *}(\hat x \hat y) < \vec x
\vec y \vert V^{(1)} \vert\phi>    =   \frac{1}{2 \sqrt{2} \pi} \sum
_{\lambda_{0}} i^{\lambda_{0}} \hat\lambda_{0}  j_{\lambda_{0}} ( q_{0}
y) u_{n_0}(x)\\
&  & \sum_{k} \hat k  \sqrt{ \hat l_{0}} (-)^{\lambda_{0} +L  +l}
\left\{ \begin{array}{ccc}
           l & \lambda & L \\
           \lambda_0 & l_0 & k
          \end{array}
\right\}  C(k l_0 l,00) C(k \lambda_0 \lambda,00) C(l_0 \lambda_0 L,m_0,0)\nonumber\\
&  & \delta_{M,m_{0}} \int_{-1}^{1} dt P_{k}(t) V^{(1)} (x, y, t)\nonumber
\end{eqnarray}

The other pieces of that second part of the driving term  can be worked out
 similarily and we obtain altogether 
\begin{eqnarray}
T_{l \lambda L}^{0,2}(x y)&  = &  \int_0^{\infty} d x' x'^2 \int_0^{\infty} d y' y'^2 
[ \frac{\delta(x-x')}{x x'}  \frac{\delta(y-y')}{y y'} \label{119}\\
&  + &  \frac{2}{\pi } \int_0^{\infty} d q q^2 j_{\lambda} ( qy) j_{\lambda} ( qy')
 \int_0^{\infty} d x'' x''^2 \tau_l( x x'';\epsilon_q) g_l(x'' x';\epsilon_q) ]\nonumber\\
& &  \frac{1}{\sqrt{2}} \sum_{\lambda_0} i^{\lambda_0} \sqrt{ \hat \lambda_0} \sum_k \hat k
 C( l_0 \lambda_0 L, m_0 0) \int_{-1}^{1} dt P_k(t) V^{(1)} ( x'y't)\nonumber\\
& &  [ \frac{1}{2 \sqrt{2}}  j_{\lambda_0} ( q_0y') u_{n_0}(x') \sqrt{\hat l_0 \hat \lambda_0} 
(-)^{\lambda_0 + L + l}
\left\{ \begin{array}{ccc}
           l & \lambda & L \\
           \lambda_0 & l_0 & k
          \end{array}
\right\}  C( k l_0 l, 00) C( k \lambda_0 \lambda,00)\nonumber\\
& + &  (-)^{l_0} \sum_k' \hat k' \int_{-1}^{1} P_{k'} (t)
 u_{n_0} ( \vert \frac{1}{2} \vec x' + \vec y' \vert )
 j_{\lambda_0} ( q_0 \vert \frac{3}{4} \vec x' -  \frac{1}{2}\vec y' \vert)\nonumber\\
& & \sum_{\mu} \frac{(-)^{\mu}}{\sqrt{\hat \mu}} C( k k' \mu, 00)^2
 \Lambda_{ l' \lambda', l_0 \lambda_0}^{ \mu l} ( x'y')
( 1 + (-)^{\lambda_0 + \lambda' + \mu + k'})]\nonumber
\end{eqnarray} 
 The remaining parts of the kernel in  Eq(\ref{41}) or (\ref{74b}) can be worked out
 similarily and we split it into two pieces:
\begin{eqnarray}
K_{l \lambda L}^{(2,1)} ( xy) & \equiv &  \int d \hat x \int d \hat y 
 \mathcal{Y}_{l \lambda}^{L M *}(\hat x \hat y)
 < \vec x \vec y \vert V^{(1)} G_0 ( 1 + \mathcal{P}) T> \label{120}\\
& = &  \frac{1}{\pi} \sum_k \hat k ( -)^k \int_{-1}^{1} dt V^{(1)}( xy,t) \sum_{\mu_2}
 \int_0^{\infty} dq q^2 j_{\mu_2} ( qy)\nonumber\\
& &  \sum_{\mu_1} \sqrt{ \hat \mu_1 \hat \mu_2} (-)^{ \mu_2 + L + k + l}
 \left\{ \begin{array}{ccc}
           l & \lambda & L \\
           \mu_2 & L & k
          \end{array}\right\}
C( k \mu_1 l,00) C( k \mu_2 \lambda,00)\nonumber\\
& &  \int_0^{\infty} dx' x'^2 \int_0^{\infty} dy' y'^2 j_{ \mu_2} ( q y') g_{\mu_1}(x x';\epsilon_q)
 [ T_{ \mu_1 \mu_2 L} ( x'y')\nonumber\\
&  + &  \sum_{k'} ( 2 \pi) \sqrt{ \hat k'} (-)^{k'} \sum_{ l' \lambda'} 
\int_{-1}^{1} dt P_{k'}(t) T_{ l' \lambda' L} ( \vert \frac{1}{2}\vec x' + \vec y' \vert,
\vert \frac{3}{4}\vec x' - \frac{1}{2}\vec y' \vert )\nonumber\\
& &  ( (-)^{l'} + (-)^{ l' + \lambda' + \mu_2} )
 \Lambda_{ \mu_1 \mu_2, l' \lambda'}^{ k' L} ( x'y') ]\nonumber
\end{eqnarray}  

and 
\begin{eqnarray}
K_{l \lambda L}^{(2,2)} ( xy) & \equiv &  \int d \hat x \int d \hat y
 \mathcal{Y}_{l \lambda}^{L M *}(\hat x \hat y) < \vec x \vec y \vert 
 t_1 G_0 V^{(1)} G_0 ( 1 + \mathcal{P}) T> \label{121}\\
& = & \frac{2}{\pi} \int_0^{\infty} dq q^2  j_{\lambda}(qy) \int_0^{\infty} dy' y'^2 j_{ \lambda} ( q y')
\int_0^{\infty} dx' x'^2 dx'' x''^2 \nonumber\\
& & \tau_l( x x'';\epsilon_q)  g_l (x'' x';\epsilon_q) K_{l \lambda L}^{(2,1)}( x'y')\nonumber
\end{eqnarray}

Then we end up with  the one integral equation for $T_{l \lambda L}(x y) $ in the form
\begin{eqnarray}
T_{l \lambda L} ( xy) = T_{ l \lambda L}^{0,1}(xy) + T_{ l \lambda L}^{0,2}(xy)
+ K_{l \lambda L}^{(1)} ( xy) + K_{l \lambda L}^{(2,1)} ( xy)+
 K_{l \lambda L}^{(2,2)} ( xy) \label{122}
\end{eqnarray}
where the $T^{0,i}$ are the driving terms and the $ K$'s incorporate the $ T$-amplitudes.

 These are coupled equations among the $T$'s, since Eqs (\ref{112}),(\ref{120})  and (\ref{121})
 contain sums over several $T$'s.
Once the T-amplitudes  in case of distinguishable particles are obtained we
 can use the Eqs. (\ref{26}),(\ref{27}),(\ref{28}) to determine the
elastic and arrangement amplitudes as well as the complete break-up amplitude (\ref{20}).
In the case of identical atoms one can not distinguish, of course, between
elastic and arrangement amplitudes. Let us first regard the elastic amplitude
given in Eq(\ref{43}).

For $< \phi^{\prime}\vert \mathcal{P}T \vert\phi>$ one expands the plane wave,
introduces the expression (\ref{113}) for T, total angular momentum states and
finally expands $u_{n^{\prime}} ( \vert\frac{4}{3}\vec y^{\prime}+ \frac{2}{3}
\vec y \vert) T_{l \lambda L} ( \vert\frac{2}{3}\vec y^{\prime}+ \frac{4}{3}
\vec y \vert,y^{\prime}) $ into Legendre polynomials. Then the same expression
(\ref{104}) for the remaining angular integrations appears and we obtain
\begin{eqnarray}
<\phi^{\prime}\vert \mathcal{P}T \vert\phi>  &  = &  2 \sqrt{2 \pi} (\frac{4}{3} )^{3}
\sum_{\lambda^{\prime}} (-i)^{\lambda^{\prime}} Y_{\lambda^{\prime}
m_{0}-m^{\prime}}(\hat q^{\prime}) \label{123}\\
&  &  \int_0^{\infty} dy y^{2} \int_0^{\infty} dy^{\prime} y'^2 \sum_{L} C(l^{\prime}
\lambda^{\prime}L,m^{\prime}m_{0}-m^{\prime}) \sum_{k} \sqrt{\hat k} (-)^{k}\nonumber\\
&   & \sum_{l \lambda} h_{l^{\prime}\lambda^{\prime},l \lambda}^{k,L} (
y,y^{\prime}) ( (-)^{l} + (-)^{l^{\prime}})\nonumber\\
&   & \int_{-1}^{1} dt P_{k}(t) u_{n^{\prime}}( \vert\frac{4}{3}\vec y^{\prime}+ \frac
{2}{3} \vec y \vert) T_{l \lambda L} ( \vert\frac{2}{3}\vec y^{\prime}+
\frac{4}{3} \vec y \vert,y^{\prime})\nonumber
\end{eqnarray}
Here $ m'$ is the magnetic quantum number of the final 2-atom bound state with orbital
 angular momentum $ l'$ and $h_{\cdots}^{\cdots}$ has been defined previously in (\ref{104}).

Very similar steps applied to $< \phi^{\prime}\vert \mathcal{P} G_{0}^{-1} \vert\phi>$
lead to
\begin{eqnarray}
<\phi^{\prime}\vert \mathcal{P} G_{0}^{-1}\vert\phi>  &  = & \frac{2}{\sqrt{\pi}}
(\frac{4}{3} )^{3} \sum_{\lambda^{\prime}} (-i)^{\lambda^{\prime}}
Y_{\lambda^{\prime}m_{0}-m^{\prime}}(\hat q^{\prime}) \label{124}\\
&   & \int dy y^{2} \int dy^{\prime}y'^2 \sum_{\lambda_{0}}
i^{\lambda_{0}} j_{\lambda^{\prime}} ( q^{\prime}y) j_{\lambda_{0}} ( q_{0}
y^{\prime}) \sqrt{\hat\lambda_{0}} ( (-)^{l_{0}} + (-)^{l^{\prime}})\nonumber\\
&  &  \sum_{L} C(l^{\prime}\lambda^{\prime}L, m^{\prime}m_{0}-m^{\prime}) C(
l_{0} \lambda_{0} L,m_0,0) \sum_{k} \sqrt{\hat k} (-)^{k} h_{
l^{\prime}\lambda^{\prime}, l \lambda}^{k,L} ( y,y^{\prime})\nonumber\\
&  &  \int dt P_{k}(t) u_{n^{\prime}}( \vert\frac{4}{3}\vec y^{\prime}+ \frac
{2}{3} \vec y \vert) V( \vert\frac{2}{3}\vec y^{\prime}+ \frac{4}{3} \vec y
\vert)  u_{n_{0}} ( \vert\frac{2}{3}\vec y^{\prime}+ \frac{4}{3} \vec y \vert)\nonumber
\end{eqnarray}

There are two more terms in (\ref{43}) which are worked out by similar steps as for
the kernel parts. The resulting expressions are
\begin{eqnarray}
< \phi' \vert V^{(1)} ( 1+\mathcal{P}) \vert \phi> &  = & 
\frac{(-)^{l_0}}{\sqrt{2}} \sum_{\lambda'} ( -i)^{\lambda'}
 Y_{\lambda' m_0-m'}(\hat q')\int_0^{\infty} dx x^2 \int_0^{\infty} dy y^2 u_{n'}(x) j_{\lambda'} ( q' y)
\label{125}\\
& &  \sum_{\lambda_0} i^{\lambda_0}
 \sqrt{\lambda_0} \sum_L C( l_0 \lambda_0 L, m_0 0) C( l' \lambda' L, m' m_0 - m')\nonumber\\ 
& &  \sum_k \hat k \int_{-1}^{1} dt P_k(t) V^{(1)} ( xyt)\nonumber\\
& &  [ \frac{1}{2 \pi} \sqrt{\hat l_0}
 (-)^{l'} \sqrt{\lambda_0} j_{\lambda_0}(q_0 y) u_{n_0} (x) (-)^L
\left\{ \begin{array}{ccc}
           l' & \lambda' & L \\
           \lambda_0 & l_0 & k
          \end{array}\right\}
C(k l_0 l',00) C(k \lambda_0 \lambda',00)\nonumber\\
&  + &  \sum_{k'} \hat k' (-)^{k'}
 \int_{1}^{1} dt P_{k'}(t) u_{n_0}( \vert \frac{1}{2} \vec x + \vec y \vert ) 
j_{\lambda_0} ( q_0 (\vert \frac{3}{4} \vec x - \frac{1}{2} \vec y \vert))\nonumber\\
& & \sum_{\mu} \frac{1}{\sqrt{\hat \mu}} C(k k' \mu,00)^2 
\Lambda_{ l' \lambda',l_0 \lambda_0 }^{\mu L} ( xy) 
( 1 + (-)^{ \lambda_0 + k + \lambda'} )]\nonumber 
\end{eqnarray}
 \begin{eqnarray}
<\phi' \vert & V^{(1)} &  ( 1+\mathcal{P}) G_0 T \vert \phi> = \frac{\sqrt{2}}{ (\pi)^{3/2}}
 \int_0^{\infty} dx x^2 \int_0^{\infty} dy y^2 \label{126} \\
& & u_{n'}(x) \sum_{\lambda'} ( -i)^{\lambda'} j_{\lambda'} ( q' y) Y_{\lambda' m_0-m'} ( \hat q')
 \sum_L C( l' \lambda' L,m' m_) - m')\nonumber\\
& & \sum_k \hat k ( -)^k \int_{-1}^{1} dt P_{k} (t)  V^{(1)}( xy,t) \sum_{\mu_2} \int_0^{\infty} dq q^2 j_{\mu_2} ( qy)\nonumber\\
& &  \sum_{\mu_1} \sqrt{ \hat \mu_1 \hat \mu_2} (-)^{ \mu_2 + L + k + l}
 \left\{ \begin{array}{ccc}
           l & \lambda & L \\
           \mu_2 & L & k
          \end{array}\right\}
C( k \mu_1 l,00) C( k \mu_2 \lambda,00)\nonumber\\
& &  \int_0^{\infty} dx' x'^2 \int_0^{\infty} dy' y'^2 j_{ \mu_2} ( q y') g_{\mu_1}(x x';\epsilon_q)
 [ T_{ \mu_1 \mu_2 L} ( x'y')\nonumber\\
&  + &  \sum_{k'} ( 2 \pi) \sqrt{ \hat k'} (-)^{k'} \sum_{ l' \lambda'}
\int_{-1}^{1} dt P_{k'}(t) T_{ l' \lambda' L} ( \vert \frac{1}{2}\vec x' + \vec y' \vert,
\vert \frac{3}{4}\vec x' - \frac{1}{2}\vec y' \vert )\nonumber\\
& &  ( (-)^{l'} + (-)^{ l' + \lambda' + \mu_2} )
 \Lambda_{ \mu_1 \mu_2, l' \lambda'}^{ k' L} ( x'y') ]\nonumber
\end{eqnarray}
Finally the complete break-up amplitude  $ < \phi_0 \vert ( 1 + \mathcal{P}) T > $ according to Eq. (\ref{42})  can be based on its first term
\begin{eqnarray}
T(\vec p \vec q )  &  \equiv & < \phi_0 \vert T>  \equiv   < \vec p \vec q \vert T>\\
&  = &  \frac{1}{(2 \pi)^{3/2}}\int d \vec x d \vec y 
e^{-i \vec p \cdot\vec x} e^{-i \vec q \cdot\vec y} 
 \sum_{l^{\prime}\lambda^{\prime}L^{\prime}} \mathcal{Y}_{l^{\prime}
\lambda^{\prime}}^{L^{\prime}m_{0}} ( \hat x \hat y) T_{l^{\prime}
\lambda^{\prime}L^{\prime}}(xy)\nonumber
\end{eqnarray}

Expanding the plane waves one readily obtains
\begin{eqnarray}
T(\vec p \vec q)  &  = &  \frac{2}{\pi}\sum_{l \lambda L} \mathcal{Y}_{l\lambda
}^{Lm_{0}} ( \hat p \hat q) (-i)^{l + \lambda}\\
& &  \int_0^{\infty} dx x^{2} \int_0^{\infty} dy y^{2} j_{l}(px) j_{\lambda}(qy) T_{l \lambda L}(xy)\nonumber
\end{eqnarray}

The remaining two terms in (\ref{42}) simply require to change the values of $\vec p $
and $\vec q$ as given in one example in( \ref{82}).

This concludes the display of the formalism. The cases for two identical atoms
out of the three or three distinguishable atoms are somewhat more tedious to
be worked out, but straightforward and are left to the interested practioner.

\bigskip

\section{The S-Wave Expressions}

For very low energy of the incoming atom and the molecule in its ground state
one can restrict the treatment to setting all orbital angular momenta to zero.
Then the final expressions simplify and allow an overview of  the
structure of the integral equation to be solved and the quadrature expression
for elastic scattering. Again we display only the case of identical atoms. In
this case the quantities $\Lambda$ and $h$ given in the Appendices both
simplify to $\frac{1}{4 \pi} $ and the integral equation (\ref{122}) composed
 of the parts(\ref{111}), (\ref{119}), (\ref{112}),(\ref{120}) and (\ref{121}) now  reads

\begin{eqnarray}
T_{000} (x,y)  &  = &  \frac{\sqrt{2}}{\pi^2} (\frac{4}{3})^3 \int_0^{\infty} dq q^{2}
j_{0}(qy) \int_0^{\infty} dy^{\prime}y'^2 j_{0}(q y^{\prime}) \int_0^{\infty} dy'' y''^2
  j_{0} ( q_{0} y'')\\
& &  \int_{-1}^{1} dt~ r_{0} ( x, \vert\frac{4}{3} \vec y''+ \frac{2}{3} \vec y^{\prime}\vert;
\epsilon_{q}) u_{n_{0}} ( \vert\frac{2}{3} \vec y^{\prime\prime}+ \frac{4}{3}
\vec y^{\prime}\vert)\nonumber\\
&  + &  \frac{1}{\sqrt{2} \pi} V(x) \int_{-1}^{1} dt u_{n_{0}} ( \vert\frac{1}{2} \vec x +
\vec y \vert) j_{0} ( q_{0} \vert\frac{3}{4} \vec x - \frac{1}{2} \vec y
\vert)\nonumber\\
& + &  \int_0^{\infty} d x' x'^2 \int_0^{\infty} d y' y'^2
[ \frac{\delta(x-x')}{x x'}  \frac{\delta(y-y')}{y y'}\nonumber\\
&  + &  \frac{2}{\pi } \int_0^{\infty} d q q^2 j_0 ( qy) j_0( qy')
 \int_0^{\infty} d x'' x''^2 \tau_0( x x'';\epsilon_q) g_0(x'' x';\epsilon_q) ]\nonumber\\
& &  \frac{1}{\sqrt{2}}  \int_{-1}^{1} dt  V^{(1)} ( x'y't) 
  [ \frac{1}{2 \sqrt{2}}  j_0 ( q_0y') u_{n_0}(x') + 
   \int_{-1}^{1} dt u_{n_0} ( \vert \frac{1}{2} \vec x' + \vec y' \vert )
 j_0 ( q_0 \vert \frac{3}{4} \vec x' -  \frac{1}{2}\vec y' \vert) \frac{1}{2 \pi}]\nonumber\\
&  +  & \frac{2}{\pi} \int_0^{\infty} dq q^{2} j_{0}(qy) \int_0^{\infty} dx''
x''^2 \int_0^{\infty} dx' x'^2 \tau_{0}( x,x^{\prime
};\epsilon_{q}) g_{0} (x^{\prime},x''; \epsilon_{q})\nonumber\\
&   & \int_0^{\infty} dy''y''^2 j_{0}(q y'') \int_{-1}^{1}
dt T_{000} ( \vert\frac{1}{2} \vec x''+ \vec y'' \vert,
 \vert\frac{3}{4} \vec x''- \frac{1}{2} \vec
y'' \vert)\nonumber\\
& + &  \frac{1}{\pi}  \int_{-1}^{1} dt V^{(1)}( xy,t)  \int_0^{\infty} dq q^2 j_0 ( qy) 
 \int_0^{\infty} dy' y'^2 j_0 ( q y') \int_0^{\infty} dx' x'^2  g_0(x x';\epsilon_q)\nonumber\\
& &  [ T_{000} ( x'y') +     \int_{-1}^{1} dt  T_{000} ( \vert \frac{1}{2}\vec x' + \vec y' \vert,
\vert \frac{3}{4}\vec x' - \frac{1}{2}\vec y' \vert )]\nonumber\\
&  + & \frac{2}{\pi} \int_0^{\infty} dq q^2  j_0(qy) \int_0^{\infty} dy' y'^2 j_0(qy')
 \int_0^{\infty}
 dx' x'^2 \_0^{\infty} dx'' x''^2 
 \tau_0( x x'';\epsilon_q)  g_0 (x'' x';\epsilon_q) \nonumber\\
& & \frac{1}{\pi}  \int_{-1}^{1} dt V^{(1)}( x'y',t)  \int_0^{\infty} dq' q'^2 j_0 ( q'y')
  \int_0^{\infty} dx''' x'''^2 \int_0^{\infty} dy''' y'''^2 j_0 ( q y''') g_0(x' x''';\epsilon_{q'})\nonumber\\
& &  [ T_{000} ( x'''y''') +     \int dt  T_{000} ( \vert \frac{1}{2}\vec x''' + \vec y''' \vert,
\vert \frac{3}{4}\vec x''' - \frac{1}{2}\vec y''' \vert )]\nonumber
\end{eqnarray}

The elastic amplitude given by the expressions (\ref{120}), ( \ref{121}),(\ref{122}), (\ref{123})
 also simplifies and one obtains
\begin{eqnarray}
< \phi' \vert U \vert \phi>  &  = &  \frac{1}{2 \pi^{2}} \int_0^{\infty} dy y^2
 \int_0^{\infty} dy' y'^2 
j_{\lambda'} (q' y) j_{0}(q_0 y')\\
&   & \int_{-1}^{1} dt u_{n_{0}} ( \vert\frac{4}{3} \vec y^{\prime}+ \frac{2}{3} \vec y
\vert) V( \vert\frac{2}{3} \vec y^{\prime}+ \frac{4}{3} \vec y \vert)
u_{n_{0}} ( \vert\frac{2}{3} \vec y^{\prime}+ \frac{4}{3} \vec y\vert)\nonumber\\
&  +  & \frac{\sqrt{2}}{2 \pi} ( \frac{4}{3})^{3} \int_0^{\infty} dy y^{2} \int_0^{\infty}
 dy' y'^2 \int_{1}^{1} dt u_{n_{0}} ( \vert\frac{4}{3} \vec y^{\prime}+ \frac
{2}{3} \vec y \vert) T_{000} ( \vert \frac{2}{3} \vec y'+ \frac{4}{3}
\vec y \vert, y')\nonumber\\
&  + & \frac{1}{2} \frac{1}{(2 \pi)^{3/2}} \int_0^{\infty} dx x^2 \int_0^{\infty} dy y^2 u_{n'}(x)
 j_0 ( q' y)  \int_{-1}^{1} dt  V^{(1)} ( xyt)\nonumber\\
& &   [   j_0(q_0 y) u_{n_0} (x)   \int_{-1}^{1} dt u_{n_0}( \vert \frac{1}{2} \vec x + \vec y \vert )
j_0 ( q_0 (\vert \frac{3}{4} \vec x - \frac{1}{2} \vec y \vert)]\nonumber\\
& +  & \frac{1}{\sqrt{2}} \frac{1}{\pi^2} \int_0^{\infty} dx x^2 \int_)^{\infty} dy y^2 
  u_{n'}(x)  j_0 ( q' y) \frac{1}{\sqrt{4 \pi}}   \int_{1}^{1} dt V^{(1)}( xy,t) 
 \int_0^{\infty} dq q^2 j_0( qy)\nonumber\\
& &   \int_0^{\infty} dx' x'^2 \int_0^{\infty} dy' y'^2 j_0 ( q y') g_0(x x';\epsilon_q)\nonumber\\
& &  [ T_{ 000} ( x'y')  +    \int_{-1}^{1} dt  T_{000} ( \vert \frac{1}{2}\vec x' + \vec y' \vert,
\vert \frac{3}{4}\vec x' - \frac{1}{2}\vec y' \vert )]\nonumber
\end{eqnarray}

At this low energy complete break up will not be  possible.

\section{Computational considerations}

The numerical solution of the integral equations for the T-amplitudes requires
 a large amount of computer time, depending on the number of angular momenta that
 enter in the  partial wave expansion, and also on the number of mesh points that
 are needed for the discretization of the integration kernels, or equivalently,
 on the number of mesh points needed for the discretization of the coordinates x and y.
We expect to minimize the number of required mesh points by using a recently developed
 spectral integral equation method (S-IEM) \cite{Cise} for solving the two-body Lippmann-Schwinger integral equation
 in configuration space that is very economical in the number of mesh points required
 for a given accuracy. This feature has now been demonstrated in several applications
 that are summarized in \cite{Cise}. For example \cite{raw},
 for the case of the binding energy
 of the He-He dimer which is very small, the calculation of the bound state wave function
 has to be carried out to large distances. For a distance of 3000 a.u. the S-IEM required
 only 200 mesh points to obtain an accuracy of three significant figures for the
 binding energy, and with 320 mesh points the accuracy increased to six significant
 figures \cite{raw}. The method consists in dividing the radial interval into partitions,
 each partition receiving a fixed number of mesh points, expanding the unknown
 wave function into a series of Chebyshev polynomials in each partition, and then 
solving for the coefficients of the expansion. Such expansions are by themselves 
very efficient, and their accuracy properties are known. In addition, the size of
 the partitions is made automatically small in the region where the wave function
 changes rapidly, and large where it changes slowly, a feature that further  contributes to
 the economy.  An additional  advantage of the spectral method is that conventional
 interpolation methods can be avoided. In the Faddeev scheme the Jacobi coordinates
 in one arrangement have to be translated into the corresponding coordinates in
 another arrangement. These translated points do not fall onto the predetermined
 mesh of points in another arrangement, hence interpolations are usually required
 with methods using fixed mesh points. The spectral method avoids this problem,
 since the Chebyshev polynomials, being analytic functions, can be evaluated at
 any prescribed positions, and hence provide the necessary translations.

The two-body $ \tau$ -  matrices play a large role in the calculation of the
 driving terms (\ref{111}) ( \ref{119})  and the integration kernels 
(\ref{112} ) ( \ref{120}) ( \ref{121}). Investigations 
 in progress are showing that the S-IEM can also be applied
 to the evaluation of the two-variable $ \tau$-  matrices, and the partition structure
 of the radial intervals of the latter will then subsequently determine the partition
 structure of the integration kernels. The number of mesh points for each of the
 two variables in the $\tau$ -matrix is expected to be of the order of 50.  The size
 of the matrices that represent the integration kernels will have to be estimated,
 and depending on the outcome, iterative procedures may be required. 

\bigskip

\section{Summary and conclusions}

The purpose of this paper is to lay out the theoretical formulation of the solution
 of the three-body Faddeev integral equations in configuration space, for the purpose
 of
 applying them to atomic physics situations. It is assumed that the Born-Oppenheimer
 approximation is valid and the three atoms are moving on  {\em one}  
 potential surface.
 The case of a conical intersection is not treated but it appears conceivable that
 also such a case could be handled by using coupled Faddeev equations, similar
 to the treatment of $ \Delta$ - excitations in a three-nucleon problem.
 We have chosen a formulation of the Faddeev equations which turned out to be 
extremely useful in nuclear physics \cite{gloeckle96}. Instead of working with
 wave function components we introduce break-up amplitudes $T_i$, which vanish 
at large distances and fulfill a set of three coupled equations. In case of identical
 atoms only one equation is needed. That set of equations is then displayed in a
 configuration space vector representation and furtheron decomposed into partial waves.
 For the sake of clarity this set of coupled equations is also presented
 assuming s- waves only. Once the T-amplitudes are determined all physical matrix elements
 for elastic and inelastic atom -diatom scattering, for the arrangement processes
 and the complete 3-atom break- up process are obtained  by simple quadrature.
 The relevant expressions are given in detailed form.

 One reason for working in configuration space is that the atom-atom two- and three-body
 potential surfaces are given in that
 space. One objection commonly raised against performing three - body calculations 
in configuration space is that one needs many mesh points for the numerical 
implementation, because the wave functions have to be calculated out to large
 distances in order to impose the asymptotic boundary conditions. We overcome this
 objection for several reasons. One is, as already said, that we do not calculate wave functions,
 but rather T-amplitudes. These functions are essentially the product of wave
 functions times potentials, and hence decay fast with distance in contrast to  wave
 functions, which oscillate at infinity. Also, the asymptotic boundary conditions of the underlying
 wave functions are automatically included, because the Faddeev  integral
 equations contain the Greens functions which lead to the appropriate asymptotic
 behavior. Another objection against working in configuration space is that wave 
functions, and so also T-amplitudes, can have a strongly oscillatory behavior,
 and hence many mesh points may be required. We expect to overcome this objection
 by using a recently developed spectral method (S-IEM) for solving the two-body
 Lippmann-Schwinger integral equation that is very economical in the number
 of mesh points required for a given accuracy \cite{Cise}.
 The economy in mesh points is crucial when solving the Faddeev integral or
 differential equations, because the numerical  complexity increases like the
 cube of the dimension of the final matrix, and the size of the dimension is
 proportional to the number of mesh points.

In summary in this paper we lay out in detail the partial wave expanded
 Faddeev equations for the $ T$-amplitudes in configuration space and in integral form.
 These equations, although complicated, are not much more complicated than the
 corresponding equations in momentum space, which have been successfully solved in the realm of nuclear physics.
 In a future study we
 intend to solve the equations for a simple test case in 
order to study the numerical feasibilty of the method. 
  
\acknowledgments
One of the authors( W.G.) would like to express his appreciation
for a Guest Professorship awarded by the  University  of Connecticut
for the period of October to December 2005. He would also like to  thankfully acknowledge
the kind hospitality extended to him  during his visit.
Both of us also would like to thank various members of the physics department,
especially  R. Cote, P. L. Gould,  H. Michels, I. Simbotin,
 W. W. Smith, and W. C. Stwally for illuminating discussions.

\appendix

\section{ The expression$h_{l\lambda,l^{\prime}\lambda^{\prime}}^{ k,L} ( y,y^{\prime})$
from Eq(\ref{104})}

The evaluation of h can be done using standard angular momentum algebra
\cite{rose} \cite{edmonds},\cite{gloeckle83}. So we do not provide the detailed steps but mention
only some useful formula, which are needed and possibly not so commonly in use.

The dependence on the two vector directions in the shifted arguments for the
spherical harmonics can be seperated using
\begin{eqnarray}
Y_{lm} ( \widehat{ \vec a + \vec b}) = \sum_{l_{1} + l_{2} = l} \frac{a^{l_{1}} b^{l_{2}
}}{\vert\vec a + \vec b \vert^{l}} \sqrt{\frac{4 \pi( 2l+1)!}{( 2l_{1} +1)! (
2l_{2} + 1)!}} \mathcal{Y}_{l_{1} l_{2}}^{lm}(\hat a,\hat b)
\end{eqnarray}

Further in the course of recoupling the following formula is needed
\begin{eqnarray}
\mathcal{Y}_{l_{1} l_{2}}^{lm} ( \hat a,\hat a) = \sqrt{ \frac{(2l_{1}+1) (
2l_{2}+1)}{4 \pi( 2l+1)}} C(l_{1} l_{2} l,00)Y_{lm}(\hat a)
\end{eqnarray}

The resulting expression for h is
\begin{eqnarray}
h_{l\lambda,l^{\prime}\lambda^{\prime}}^{k,L} ( y,y')  & \equiv &
\int d \hat y  \int d \hat y' \mathcal{Y}_{l,\lambda}^{\ast L,M}
( \frac{4}{3} \vec y'+ \frac{2}{3} \vec y , \hat y')
  \mathcal{Y}_{k,k}^{0,0}( \hat y, \hat y') 
 \mathcal{Y}_{l_{0},\lambda_{0}}^{ L ,m_0}
(  \frac{2}{3} \vec y' + \frac{4}{3} \vec y , \hat y')\\
& = &  \frac{1}{8
\pi} \sum_{l_{1} + l_{2} = l} \sum_{l_{1}^{\prime}+ l_{2}^{\prime}  =   l^{\prime
}} y^{l_{1} + l_{1}^{\prime}} {y^{\prime}}^{l_{2} + l_{2}^{\prime}} ( \frac
{2}{3})^{l_{1} + l_{2}^{\prime}} ( \frac{4}{3})^{l_{2} + l_{1}^{\prime}}\\
&   & \sqrt{\frac{( 2l+1)!}{( 2l_{1} +1)! ( 2l_{2} + 1)!}} \sqrt{ \frac{(
2l^{\prime}+1)!}{( 2l_{1}^{\prime}+1)! ( 2l_{2}^{\prime}+ 1)!}}\nonumber\\
&   & \sqrt{\hat l \hat l_{1} \hat\lambda} \sqrt{\hat l^{\prime}\hat
l_{2}^{\prime}\hat\lambda^{\prime}} \sqrt{ \hat k \hat l_{1}^{\prime}}\nonumber\\
&   & \sum_{f} 
\left\{ \begin{array}{ccc}
           l_2 & l_1 & l \\
           \lambda & L & f
          \end{array}
\right\} 
 C(l_{1} \lambda f,00) \sum_{f^{\prime}}
\left\{ \begin{array}{ccc}
           l_1' & l_2' & l' \\
           \lambda' & L & f'
          \end{array}
\right\}
C(l_{2}^{\prime},\lambda^{\prime}f^{\prime},00) \sum_{h} \hat h (-)^{h}\nonumber\\
&  &  \sum_{\gamma_{1}} ( -)^{\gamma_{1} + k}
\left\{ \begin{array}{ccc}
           f & l_2 & L \\
           f' & l_1' & \gamma_1
          \end{array}
\right\} 
  C(kh\gamma_{1},00)^{2}
C(\gamma_{1} l_{1}^{\prime}f,00) C(\gamma_{1} f^{\prime}l_{2},00)\nonumber\\
&   & \int_{-1}^{1} dt P_{h}(t) \frac{1}{\vert\frac{4}{3} \vec y^{\prime}+ \frac{2}{3}
\vec y \vert^{l}} \frac{1}{\vert\frac{2}{3} \vec y^{\prime}+ \frac{4}{3} \vec
y \vert^{l^{\prime}}}\nonumber
\end{eqnarray}

\section { The expression $\Lambda_{l\lambda,l' \lambda'}^{k,L}(xy) $ from
Eq(\ref{109})}
\begin{eqnarray}
\Lambda_{l\lambda,l^{\prime}\lambda^{\prime}}^{k,L}(xy) & \equiv & 
 \int d \hat x  d \hat y  \mathcal{Y}_{l,\lambda}^{\ast\ L,M}( \hat x, \hat y )
\mathcal{Y}_{k,k}^{\ 0,0}( \hat x ,\hat y ) \mathcal{Y}_{l_0 \lambda_ 0 }^{L m_0 }
(  \frac{1}{2} \vec x + \vec y, \frac{3}{4} \vec x - \frac{1}{2} \vec y)\\
 &  = & \frac{1}{8\pi}
\sum_{l_{1}^{\prime}+l_{2}^{\prime}=l^{\prime}}\sum_{\lambda_{1}^{\prime
}+\lambda_{2}^{\prime}=\lambda^{\prime}}(\frac{1}{2})^{l_{1}^{\prime}
+\lambda_{2}^{\prime}}(\frac{3}{4})^{\lambda_{1}^{\prime}}(-)^{\lambda
_{2}^{\prime}}x^{l_{1}^{\prime}+\lambda_{1}^{\prime}}y^{l_{2}^{\prime}
+\lambda_{2}^{\prime}}\\
&   & \sqrt{\frac{(2l^{\prime}+1)!}{(2l_{1}^{\prime})!(2l_{2}^{\prime}!}}
\sqrt{\frac{(2\lambda^{\prime}+1)!}{(2\lambda_{1}^{\prime})!(2\lambda
_{2}^{\prime})!}}\nonumber\\
&   & \sum_{\mu_{1}\mu_{2}}\sqrt{\hat{l}^{\prime}\hat{\lambda}^{\prime}}
\left\{ \begin{array}{ccc}
           l_1' & l_2' & l' \\
           \lambda_1' & \lambda_2' & \lambda' \\
           \mu_1 & \mu_2 & L
          \end{array}
\right\} 
C(l_{1}^{\prime}\lambda_{1}^{\prime}\mu_{1},00)C(l_{2}^{\prime}\lambda
_{2}^{\prime}\mu_{2},00)\nonumber\\
&   & \sum_{h}\sqrt{h}(-)^{h}\sum_{\gamma_{1}}\sqrt{\hat{k}\hat{h}}C(kh\gamma
_{1},000)\sqrt{\hat{\mu}_{1}\hat{\mu}_{2}}(-)^{\mu_{2}+L+\gamma_{1}}
(-)^{l}
\left\{ \begin{array}{ccc}
           l & \lambda & L \\
           \mu_2 & \mu_1 & \gamma_1
          \end{array}
\right\} \nonumber\\
&   & C(\gamma_{1}\mu_{1}l,00)C(\gamma_{1}\mu_{2}\lambda,00)\nonumber\\
&   & \int_{-1}^{1} dtP_{h}(t)\frac{1}{|\frac{1}{2}\vec{x}+\vec{y}|^{l^{\prime}}}\frac
{1}{|\frac{3}{4}\vec{x}-\frac{1}{2}\vec{y}|^{\lambda^{\prime}}}\nonumber
\end{eqnarray}

\bigskip

\end{document}